\documentclass[aps,pre,preprint,amsmath,amssymb,nofootinbib]{revtex4-1}
\usepackage[percent]{overpic}

\begin{document}
\begin{flushleft}
\textit{Journal of Experimental and Theoretical Physics, 2011, Vol. 113, No. 4, pp. 592--604.}\footnote{%
\baselineskip=5pt
Original Russian Text has been published in
Zhurnal Eksperimental'noi i Teoreticheskoi Fiziki, 2011, Vol.~140, No.~4, pp.~681--695.}
\end{flushleft}
\title{Absorption of Gamma-Ray Photons\\
in a Vacuum Neutron Star Magnetosphere:\\
I. Electron-Positron Pair Production}
\author{Ya.\ N. \surname{Istomin}}
\email{\texttt{istomin@lpi.ru}}
\author{D. N. \surname{Sob'yanin}}
\email{\texttt{sobyanin@lpi.ru}}
\affiliation{Lebedev Physical Institute, Russian Academy of Sciences,\\Leninskii pr.\ 53, Moscow, 119991 Russia}
\received{December 1, 2010}
\begin{abstract}
The production of electron-positron pairs in a vacuum neutron star magnetosphere is investigated for both low (compared to the Schwinger one) and high magnetic fields. The case of a strong longitudinal electric field where the produced electrons and positrons acquire a stationary Lorentz factor in a short time is considered. The source of electron-positron pairs has been calculated with allowance made for the pair production by curvature and synchrotron photons. Synchrotron photons are shown to make a major contribution to the total pair production rate in a weak magnetic field. At the same time, the contribution from bremsstrahlung photons may be neglected. The existence of a time delay due to the finiteness of the electron and positron acceleration time leads to a great reduction in the electron-positron plasma generation rate compared to the case of a zero time delay. The effective local source of electron-positron pairs has been constructed. It can be used in the hydrodynamic equations that describe the development of a cascade after the absorption of a photon from the cosmic gamma-ray background in a neutron star magnetosphere.
\end{abstract}
\maketitle
\newpage
%%%%%%%%%%%%%%%%%%%%%%%%%%%%%%
%%%%%%%%%%%%%%%%%%%%%%%%%%%%%%
\section{INTRODUCTION}

Over more than 40-year-long history of studying radio pulsars, attention has been focused on the investigation of normal pulsars. However, there exist nonstationary radio sources associated with neutron stars. The characteristics of their radio emission vary with
time, while the radio emission itself can disappear for a while. In particular, intermittent pulsars and rotating radio transients can be classified as nonstationary
radio sources. Observational data on the first intermittent pulsar were published in 2006 \citep{KramerEtal2006}. A significant detected effect is the difference between the neutron star spin-down rates in the periods of emission and silence. Analysis of archival data from the Multibeam Pulsar Survey with the Australian Parkes radio telescope has revealed rotating radio transients (RRATs)---the sources of short and relatively bright
single radio bursts \citep{McLaughlinEtal2006}. The width of individual bursts
is $2-30$~ms and the flux density reaches $10$~Jy at $1.4$~GHz \citep{KeaneEtal2010}. For most radio pulsars, the flux density at this frequency does not exceed several mJy ($1\text{ Jy}=10^{-26}\text{ W}\,\text{m}^{-2}\,\text{Hz}^{-1}$). Note that RRATs can also be
observed at a low radio frequency of 111~MHz \citep{Shitov2009}. Although the time interval between two successive bursts is random, it is a multiple of some fixed period
that lies within the range $0.1-6.7$~s and is believed to be the neutron star rotation period. This rotation period exceeds that of typical radio pulsars, which is $\sim0.3$~s.

Generally, it is rather difficult to detect nonstationary radio sources and to determine their characteristics. One of the main reasons for this is their long stay in a state of silence. Such silence is also observed in nulling pulsars; the ratio of the nulling time to the total time of pulsar observation can reach 95\% \citep{WangEtal2007}.
However, silence itself is not a necessary condition for nonstationarity, because radio emission nonstationarity also manifests itself in ordinary radio pulsars as mode switching. Observations in recent years suggest that the spin-down rate of a pulsar also changes in the absence of its switch-off \citep{LyneEtal2010}. This change is abrupt, is
often quasi-periodic, and correlates with the observed change in pulse shape for six pulsars.

Magnetars are also characterized by a transient radio emission. Magnetars are neutron stars with superstrong surface magnetic fields of the order of or higher than the Schwinger magnetic field. Their activity manifests itself mainly in bright X-ray and gamma-ray bursts. Their radio emission is observed at both low \citep{ShitovEtal2000,MalofeevEtal2005,
MalofeevEtal2007,MalofeevEtal2010} and high \citep{CamiloEtal2006,CamiloEtal2007} radio frequencies. The radio emission could be related to X-ray bursts, but there exists a magnetar, PSR~J1622--4950, without any observable bursty behavior in the X-ray band \citep{LevinEtal2010}. Moreover, based on archival data, we can assert that this radio source can be switched off for a time of about several hundred days. Since plasma outflows from the magnetosphere are currently believed to be responsible for the observed radio emission from neutron stars (see, e.g., the review \citep{Beskin1999}), the activity of the mentioned radio sources suggests a possible cessation of the plasma generation in their magnetospheres \citep{GurevichIstomin2007}. Analysis of observations for the intermittent radio pulsars PSR B1931+24 and PSR J1832+0029 shows that the observed change in neutron star spin-down rate in the absence of radio emission can be explained only if the neutron star magnetosphere becomes a vacuum one: the plasma not only is not generated in the magnetosphere but also does not come from the neutron star surface \citep{GurevichIstomin2007}. However, the plasma production through the absorption of photons from the external cosmic gamma-ray background can occur in a vacuum magnetosphere \citep{IstominSobyanin2009,IstominSobyanin2010a,
IstominSobyanin2010b}.

Let there be a primary galactic photon whose energy and propagation direction are such that the transverse momentum component exceeds~$2m_e c$, where $m_e$ is the electron mass and $c$ is the speed of light. This photon can then produce an electron-positron pair in a magnetic field. In general, the pair particles are produced at high Landau levels and, passing to the zeroth Landau level, emit synchrotron photons. In a vacuum magnetosphere, there is a strong longitudinal electric field~$E_\parallel$ whose characteristic strength is defined by the relation
\[
E_\parallel/B\sim\Omega R_S/c=R_S/R_L\sim10^{-4},
\]
where $R_S\approx10$~km is the neutron star radius, $R_L=c/\Omega$ is the light cylinder radius, and $\Omega$ is the angular frequency of neutron star rotation. The particles in such a field acquire some stationary Lorentz factor $\gamma_0$. This Lorentz factor can be determined from the condition of balance between the work of the longitudinal electric field and the curvature radiation power and can reach $\gamma_0\sim10^7-10^8$ in a vacuum field \citep{IstominSobyanin2009,IstominSobyanin2010a}. We have curvature and synchrotron photons that, in turn, produce electron-positron pairs. The secondary particles also emit synchrotron photons, acquire the stationary Lorentz factor $\gamma_0$, and emit curvature photons, which the particles of the primary pair continue to do. Subsequently, the next generation of particles is produced and so on. To study the development of an electron-positron plasma generation cascade triggered by the absorption of a photon from the external cosmic gamma-ray background, it is necessary to calculate the source of electron-positron pairs. This is the goal of our work. Here, it should be pointed out that the problem under consideration differs significantly from the problem of stationary electron-positron plasma generation in a neutron star magnetosphere filled with a dense plasma \citep{Gurevich1985,BGI1993}. A strong longitudinal electric field in the latter case exists only in the polar region of the magnetosphere near the neutron star surface, so that the plasma is produced only in this bounded region and the generation cascade is usually limited to two generations.

The paper is structured as follows. In Section~2, we derive the integral equation for a self-consistent source of electron-positron pairs by taking into account the particle production by curvature and synchrotron photons and find its solution. In Section~3, we consider the initial stage of development of the pair production cascade when the hydrodynamic equations are not yet applicable. In Section~4, we investigate the influence of the time delay between the production of particles and the absorption of the photons emitted by them in a magnetic field on the electron-positron plasma generation rate and calculate the effective local source of electron-positron pairs. Our main conclusions are presented in Conclusions.

\section{PAIR PRODUCTION}

Consider the filling of the magnetosphere of a neutron star with a surface magnetic field $B\sim10^{12}$~G. We assume the initial state of the magnetosphere to be a vacuum one. This implies that the existing charge density in the magnetosphere is small compared to the Goldreich-Julian density $\rho_{GJ}=-\mathbf{\Omega}\cdot\mathbf{B}/2\pi c$, when the longitudinal electric field is screened. The corresponding electromagnetic field was calculated in \citep{Deutsch1955}. If we assume that there is no free charge escape from the stellar surface, then the magnetosphere can be filled only through the production of electron-positron pairs by high-energy gamma-ray photons from the external Galactic background. Allowance for the influence of these photons on the particle production processes in pulsars was shown to be necessary in~\citep{ShukreRadhakrishnan1982}. It was pointed out in the latter paper that an influx of new charges that can be provided by the absorption of photons from the external cosmic gamma-ray background is needed in the Ruderman-Sutherland model \citep{RudermanSutherland1975} to trigger each new spark in the polar gap of a pulsar magnetosphere. We will add that when the polar cap of a radio pulsar is considered, the case of free particle escape from the surface of a neutron star with a nondipolar magnetic field is also investigated \citep{BarsukovKantorTsygan2007,BarsukovPolyakovaTsygan2009}.

In what follows, we will measure the magnetic field strength in units of the Schwinger (critical) field
\begin{equation*}
\label{criticalField}
B_{cr}=\frac{m_e^2c^3}{e\hbar}\approx4.414\times10^{13}\text{ G},
\end{equation*}
where $e$ is the positron charge. As the units of mass, length, and time, we will take, respectively, the electron mass $m_e\approx9.109\times10^{-28}$~g, the Compton electron
wavelength $^-\!\!\!\!\lambda=\hbar/m_e c\approx3.862\times10^{-11}$~cm, and its ratio to the speed of light $^-\!\!\!\!\lambda/c\approx1.288\times10^{-21}$~s. Formally, this means that $\hbar={^-\!\!\!\!\lambda}=c=1$. Thus, the length, velocity, time, and energy will be measured in units of~$^-\!\!\!\!\lambda$, $c$, $^-\!\!\!\!\lambda/c$, and~$m_e c^2$, respectively. We will note at once that in these units
$1\text{ cm}\approx2.590\times10^{10}$ and $1\text{ s}\approx7.763\times10^{20}$.

We consider the production of an electron-positron plasma in a weak magnetic field $B\ll1$. In this case, the number of electron-positron pairs produced per unit time per unit volume in a unit interval of longitudinal Lorentz factors is given by the expression \citep{IstominSobyanin2007}
\begin{equation}
\label{Q3}
q_p=\frac{2a}{\Lambda}q_{ph}\Bigl(2\gamma_\parallel\frac{a}{\Lambda}\Bigr),
\end{equation}
where $q_{ph}(k)$ is the number of photons produced per unit time per unit volume in a unit energy interval, $a=4/3B\sim100$, $\gamma_\parallel$ is the longitudinal Lorentz factor of a particle, and $\Lambda\sim10\gg1$ is the logarithmic factor. We
see that the source of particles $q_p$ is proportional to the source of photons $q_{ph}$, which is represented as the sum of the source of curvature photons $q_{curv}$ and the source of synchrotron photons~$q_{syn}$:
\begin{equation}
\label{weakFieldSmallQ}
q_{ph}(k)=q_{curv}(k)+q_{syn}(k).
\end{equation}

The curvature photons are produced by ultrarelativistic particles as they move along curved magnetic field lines and have a characteristic energy
\begin{equation}
\label{curvatureEnergy}
k_{curv}=\frac{3}{2}\frac{\gamma_\parallel^3}{\rho},
\end{equation}
where $\rho$ is the radius of curvature of the particle trajectory, which virtually coincides with the radius of curvature of the magnetic field lines. The number of curvature photons produced per unit time in a unit energy interval near~$k$ is given by the expression
\begin{equation}
\label{summaryCurvatureEmissionProbability}
P_{curv}(\gamma_\parallel,k)=\frac{1}{\sqrt{3}\,\pi}\frac{\alpha}{\gamma_\parallel^2}\,\varphi\left(\frac{k}{k_{curv}}\right),
\end{equation}
where
\begin{equation}
\label{phi}
\varphi(x)=\int\limits_x^\infty K_{5/3}(y)\,dy,
\end{equation}
$K_{5/3}(y)$ is the $5/3$-order Macdonald function, $\alpha= e^2/\hbar c\approx1/137$ is the fine-structure constant.

Let us introduce the spectral particle distribution function $F(\gamma_\parallel)$ equal to the number of particles per unit volume per unit interval of longitudinal Lorentz factors. The source of curvature photons can then be represented as
\begin{equation}
\label{curvaturePhotonSource}
q_{curv}(k)=\int\limits_{\gamma_{\min}}^\infty P_{curv}(\gamma_\parallel,k)F(\gamma_\parallel)\,d\gamma_\parallel,
\end{equation}
where $\gamma_{\min}\sim100$ is the minimum Lorentz factor of the produced particles \citep{IstominSobyanin2007}.

Let us calculate the source of synchrotron photons $q_{syn}(k)$. Let initially there be a photon with an energy $k_i$ propagating at an angle $\chi$ to the magnetic field direction. An electron-positron pair whose particles initially have a Lorentz factor $\gamma_i$ and move at an angle $\theta_i$ to the magnetic field direction is produced after its absorption. Subsequently, the particles lose their transverse velocity component through synchrotron radiation and pass to the zeroth Landau level, while having some final Lorentz factor $\gamma_f=\gamma_\parallel$ that we call the longitudinal one. The law of conservation of the energy $\gamma_i=k_i/2$ \citep{Beskin1982} and the law of conservation of the longitudinal momentum component $k_i\cos\chi=2\gamma_i v_\parallel$, where $v_\parallel=v_i\cos\theta_i$ is the longitudinal electron or positron velocity component and $v_i$ is the initial particle velocity, hold at the instant of electron-positron pair production. It is important to note that the longitudinal velocity component remains unchanged during the emission of synchrotron photons: $v_\parallel=v\cos\theta$, where $v=\sqrt{1-1/\gamma^2}$ and $\theta$ are some intermediate values of the particle velocity and pitch angle, $\gamma$ is an intermediate Lorentz factor of the particle. The photons propagate at small angles to the magnetic field, $\chi\ll1$. Consequently, the pitch angles of the particles produced by them are also small, because $0\leqslant\theta\leqslant\theta_i<\chi$. Using the fact that, in this case, $\sin\theta\approx\theta$, we have the following dependence of the particle pitch angle on its current Lorentz factor:
\begin{equation}
\label{pitchAngleThroughGamma}
\theta^2=\chi^2\left(1-\frac{1}{\gamma^2\chi^2}\right).
\end{equation}
Setting the pitch angle equal to zero gives the longitudinal Lorentz factor
\begin{equation}
\label{gammaOnChiDependence}
\gamma_\parallel=1/\chi.
\end{equation}
To simplify the calculations, it is convenient to introduce an angle $\eta$ via the relation $\cos\eta=1/\gamma\chi$. As we see from Eq.~\eqref{pitchAngleThroughGamma}, the pitch angle is then given by the expression
\begin{equation}
\label{pitchAngleThroughEta}
\theta=\chi\sin\eta,
\end{equation}
with $\eta_i=\arccos(2/k_i\chi)$ corresponding to the initial pitch angle $\theta_i$.

The number of synchrotron photons emitted by a particle with a Lorentz factor $\gamma$ and a pitch angle $\theta$ per unit time in a unit energy interval near $k$ is
\begin{equation}
\label{synchrotronEmissionIntensity}
P_{syn}(\theta,\gamma,k)=\frac{1}{\sqrt{3}\pi}\frac{\alpha}{\gamma^2}\,\varphi\left(\frac{k}{k_{syn}}\right),
\end{equation}
where
\begin{equation*}
\label{characteristicSynchrotronPhotonEnergy}
k_{syn}=\frac{3}{2}B\theta\gamma^2
\end{equation*}
is the characteristic energy of synchrotron photons. During the emission of synchrotron photons, the particle energy decreases:
\begin{equation}
\label{energyDecayThroughSynchrotronEmission}
\frac{d\gamma}{dt}=-\frac{2}{3}\alpha\left(B\theta\gamma\right)^2.
\end{equation}
Here, the total intensity of the synchrotron radiation obtained by integrating Eq.~\eqref{synchrotronEmissionIntensity} with the weight $k$ over all energies from~$0$ to~$\infty$ appears on the right-hand side.

To find the total number of photons $\mathcal{N}_{syn}$ in a unit energy interval near $k$ emitted by one particle in the synchrotron radiation time, Eq.~\eqref{synchrotronEmissionIntensity} should be integrated over all times by taking into account the time dependence of the particle Lorentz factor and pitch angle. Integration over the angle $\eta$ from~$\eta_i$ to~$0$ corresponds to integration over the time from~$0$ to~$\infty$. Given Eqs.~\eqref{pitchAngleThroughGamma}, \eqref{pitchAngleThroughEta}, and~\eqref{energyDecayThroughSynchrotronEmission}, it is easy to obtain the expression
\begin{equation*}
\label{integratedSynchrotronPhotonDistributionFunction}
\mathcal{N}_{syn}\left(k_i,\chi,k\right)=\frac{\sqrt{3}}{2\pi}\frac{\chi}{B^2}\int\limits_0^{\eta_i}
d\eta\frac{\cos^2\eta}{\sin\eta}\,\varphi\left(a\frac{k\chi}{2}\frac{\cos^2\eta}{\sin\eta}\right).
\end{equation*}

The spectral distribution $\mathcal{N}_{syn}$ depends on the energy $k_i$ of the initial photon that produced an electron-positron pair and its pitch angle~$\chi$. However, these two quantities are not independent and their relation is given by the condition for photon absorption in a magnetic field \citep{IstominSobyanin2007}:
\begin{equation*}
\label{absorptionCriterion}
\frac{k_i\chi}{2}=\frac{a}{\Lambda}=\frac{1}{\cos\eta_i}.
\end{equation*}
Since the relation $a/\Lambda\sim10\gg1$ holds, the angle $\eta_i$ is
close to~$\pi/2$ or, more specifically,
\begin{equation*}
\label{angleEtaI}
\eta_i\approx\frac{\pi}{2}-\frac{\Lambda}{a}.
\end{equation*}
Given Eq.~\eqref{gammaOnChiDependence}, the source of synchrotron photons can be written as
\begin{equation}
\label{synchrotronPhotonSource}
q_{syn}(k)=2\int\limits_{\gamma_{\min}}^\infty q_p(\gamma_\parallel)\,\mathcal{N}_{syn}\left(2\gamma_\parallel\frac{a}{\Lambda},\frac{1}{\gamma_\parallel},k\right)d\gamma_\parallel.
\end{equation}

The electron-positron pair production rate $q_p$ is proportional to the total source of photons $q_{ph}$ containing the part responsible for the synchrotron photons. In turn, the synchrotron photon production rate $q_{syn}$ itself is proportional to the pair production rate $q_p$. Given Eqs.~\eqref{Q3}, \eqref{weakFieldSmallQ}, and~\eqref{synchrotronPhotonSource}, this self-consistency condition allows the integral equation to find the source of electron-positron pairs $q_p$ to be derived:
\begin{equation}
\label{integralEquationForFindingPairSource}
q_p(\gamma_\parallel)=\frac{2a}{\Lambda}q_{curv}\left(2\gamma_\parallel\frac{a}{\Lambda}\right)
+\frac{a^3}{\Lambda}\xi\int\limits_{\gamma_{\min}}^\infty
\frac{d\gamma'}{\gamma'}q_p(\gamma')\int\limits_0^{\eta_i}d\eta\frac{\cos^2\eta}{\sin\eta}
\varphi\left(\frac{a^2}{\Lambda}\frac{\gamma_\parallel}{\gamma'}\frac{\cos^2\eta}{\sin\eta}\right),
\end{equation}
where
\begin{equation*}
\label{xi}
\xi=\frac{9\sqrt3}{8\pi}.
\end{equation*}

Consider some volume of an electron-positron plasma in a magnetic field. The plasma production process can be represented as follows. Let there be some photon, a curvature or synchrotron one, that produces an electron-positron pair. After the emission of synchrotron photons, the particles at the zeroth Landau level have some longitudinal Lorentz factor and, consequently, emit curvature photons. As we see from Eq.~\eqref{curvaturePhotonSource}, the source of curvature photons depends on the spectral distribution of particles in longitudinal Lorentz factor. Below, we will consider the case where a fairly strong longitudinal electric field exists. The electron and the positron then accelerate in the time
\begin{equation*}
\tau_{st}=\frac{\gamma_0}{E_\parallel}
\end{equation*}
and move in opposite directions with the characteristic Lorentz factors
\begin{equation*}
\label{gammaMax}
\gamma_0=\left(\frac{3}{2\alpha}E_\parallel\rho^2\right)^{1/4}
\end{equation*}
(see \citep{IstominSobyanin2009,IstominSobyanin2010a}). Here, we do not assume that the field $E_\parallel$ is equal to the external electric field in a vacuum magnetosphere. This field is the total longitudinal electric field that is the sum of the external field and the plasma screening field. The field strength is assumed to be sufficient for electron-positron pairs to be efficiently produced. Thus, we can write the total particle distribution function as
\begin{equation}
\label{particleNumberDensity}
F(\gamma_\parallel)=(n_++n_-)\delta(\gamma_\parallel-\gamma_0),
\end{equation}
where $n_+$ and $n_-$ are the positron and electron number densities, respectively, and $\delta(x)$ is the delta function. This by no means implies that the source of particles $q_p(\gamma_\parallel)$ must be proportional to the delta function, because it depends on the source of photons without a monochromatic spectrum. Nevertheless, the produced particles will acquire the Lorentz factor $\gamma_0$ irrespective of the initial particle spectrum. For this reason, it is important for us to find the total number of electron-positron pairs produced per unit time per unit volume.

The particle distribution function \eqref{particleNumberDensity} is proportional to the total particle number density. Since the integral equation \eqref{integralEquationForFindingPairSource} is linear, the source of electron-positron pairs is also proportional to $n_++n_-$. Below, it is convenient to assign the function $q_p$ to one particle by formally assuming in all of the previous formulas that $n_++n_-=1$. In particular, the source of curvature photons $q_{curv}(k)$ will be given by Eq.~\eqref{summaryCurvatureEmissionProbability}, in which we should set $\gamma_\parallel=\gamma_0$. The number of electron-positron pairs produced per unit time per particle is then
\begin{equation}
\label{normalizedTotalPlasmaSource}
Q(\gamma_{\min})=\int\limits_{\gamma_{\min}}^\infty q_p(\gamma_\parallel)d\gamma_\parallel.
\end{equation}

Above, we have not mentioned that, apart from curvature and synchrotron photons, the produced particles also emit bremsstrahlung photons because of their acceleration in the longitudinal electric field~$E_\parallel$. The bremsstrahlung intensity is \citep{SokolovTernov1974}
\begin{equation*}
\label{bremsstrahlungIntensity}
W_{brems}=\frac{2}{3}\alpha E_\parallel^2.
\end{equation*}
It should be compared with the intensity of the curvature radiation
\begin{equation}
\label{curvatureRadiationIntensity}
W_{curv}=\frac{2}{3}\alpha\frac{\gamma^4}{\rho^2},
\end{equation}
where $\gamma$ is the particle Lorentz factor. When the condition of quasi-stationary motion $W_{curv}=E_\parallel$ corresponding to $\gamma=\gamma_0$ is met, we have
\begin{equation}
\label{bremsstrahlungToCurvatureIntensityRatio}
\frac{W_{brems}}{W_{curv}}=\frac{2}{3}\alpha E_\parallel\ll1.
\end{equation}
At $E_\parallel\sim10^{-4}$ this ratio is equal in order of magnitude to $W_{brems}/W_{curv}\sim10^{-6}$. Obviously, condition~\eqref{bremsstrahlungToCurvatureIntensityRatio} is always met, because the longitudinal electric field is definitely lower than the Schwinger one, $E_\parallel<1$. Otherwise, we would have the direct production of electron-positron pairs from a vacuum, which would immediately lead to electric field screening.

However, the particle initially has a Lorentz factor much smaller than~$\gamma_0$. The brems\-strahlung intensity will then still exceed the intensity of the curvature radiation until the Lorentz factor reaches $\sqrt{\rho E_\parallel}$. For $E_\parallel\sim10^{-4}$ and $\rho\sim10^{17}$, the characteristic transition Lorentz factor is~$10^6$. To estimate the total contribution from bremsstrahlung photons, let us find the total bremsstrahlung energy over the entire time of particle acceleration to the stationary Lorentz factor~$\gamma_0$. In this case, we assume that the initial particle Lorentz factor is much smaller than~$\gamma_0$. This will allow us to estimate an upper limit for the bremsstrahlung energy, because, obviously, it is proportional to the difference between the final and initial Lorentz factors of the particle being accelerated. The total energy of the emitted bremsstrahlung photons in the time $\tau_{st}$ will be
\begin{equation*}
\label{totalBremsstrahlungEnergy}
E_{brems}=W_{brems}\tau_{st}.
\end{equation*}
The total energy of the curvature radiation in the particle acceleration time is obtained by integrating Eq.~\eqref{curvatureRadiationIntensity}, in which the current Lorentz factor $\gamma_\parallel=E_\parallel t$ should be taken instead of~$\gamma$, over the time from~$0$ to~$\tau_{st}$:
\begin{equation*}
\label{totalCurvatureEnergy}
E_{curv}=W_{curv}\frac{\tau_{st}}{5}.
\end{equation*}
Now, it is easy to obtain the relation
\begin{equation*}
\label{bremsstrahlungToCurvatureEnergyRatio}
\frac{E_{brems}}{E_{curv}}=5\frac{W_{brems}}{W_{curv}}\ll1.
\end{equation*}
Thus, the ratio of the total energy of the bremsstrahlung photons to the total energy of the curvature photons emitted in the time of particle acceleration to the stationary Lorentz factor $\gamma_0$ is very small and does not depend on the value of~$\gamma_0$ itself. We will add that there is an additional contribution from synchrotron photons to the total pair production rate in a relatively weak field $B\ll1$, which, as a rule, exceeds the contribution from curvature photons. Consequently, the presence of bremsstrahlung photons may be disregarded.

\subsection{The Mellin Transform}

The integral equation \eqref{integralEquationForFindingPairSource} can be solved as follows. Consider the second term on the right-hand side of Eq.~\eqref{integralEquationForFindingPairSource}. In this double integral, we will change the order of integration and initially consider the integral
\begin{equation*}
\label{integralOfSourceOverGamma}
I=\int\limits_{\gamma_{\min}}^\infty
\frac{d\gamma'}{\gamma'}q_p(\gamma')\varphi\left(\frac{\Upsilon}{\gamma'}\right),
\end{equation*}
where
\begin{equation*}
\label{Upsilon}
\Upsilon=\frac{a^2}{\Lambda}\gamma_\parallel\frac{\cos^2\eta}{\sin\eta}.
\end{equation*}
The quantity $\Upsilon$ is limited from below by
\begin{equation*}
\label{upsilonConstriction}
\Upsilon\geqslant\Lambda\gamma_\parallel\geqslant\Lambda\gamma_{\min}\gg1
\end{equation*}
and, in any case, $\Upsilon\geqslant1000$. The integral $I$ is a repeated one (see~\eqref{phi}) and, once the order of integration has been changed, can be represented as
\begin{equation}
\label{transformedI}
I=\varphi\left(\frac{\Upsilon}{\gamma_{\min}}\right)qi_p(\gamma_{\min})
+\int\limits_0^{\Upsilon/\gamma_{\min}}dy\,K_{5/3}(y)\,qi_p\left(\frac{\Upsilon}{y}\right),
\end{equation}
where we introduced a function
\begin{equation*}
\label{qiDefinition}
qi_p(x)=\int\limits_x^\infty q_p(y)\frac{dy}{y}.
\end{equation*}
Below, we will consider the functions $q_p(x)$ and $qi_p(x)$ for all values of the argument $x$ from~$0$ to~$\infty$. We do not restrict ourselves to $x\geqslant\gamma_{\min}$ at which the function $q_p(x)$ is initially defined and assume that $q_p(x)$ was somehow extended to the interval $(0,\gamma_{\min})$. For us, it is only important that the function $q_p(x)$ coincides with the solution of the integral equation \eqref{integralEquationForFindingPairSource} for $x\equiv\gamma_\parallel\geqslant\gamma_{\min}$.

Let us take a power function as a trial one:
\begin{equation}
\label{trialQi}
qi_p(x)=x^{-s}.
\end{equation}
Here, $s$ is some complex number such that integral \eqref{transformedI} is a priori assumed to be convergent. Let us introduce the power moments of the Macdonald function $K_{5/3}(x)$ using the relations
\begin{equation*}
\label{McDonaldFunctionIncompleteMoment}
M_s(x)=\int\limits_x^\infty y^s K_{5/3}(y)\,dy,
\end{equation*}
\begin{equation*}
\label{McDonaldFunctionCompleteMoment}
M_s\equiv M_s(0)
=2^{s-1}\Gamma\left(\frac{s}{2}+\frac{4}{3}\right)\Gamma\left(\frac{s}{2}-\frac{1}{3}\right),
\end{equation*}
where $\Gamma(x)$ is Euler's gamma function \citep{GradsteinRyzhik1963}. The integral on the right-hand side of Eq.~\eqref{transformedI} will then be
\begin{equation*}
\label{McDonaldQiIntegral}
\int\limits_0^{\Upsilon/\gamma_{\min}}dy\,K_{5/3}(y)\,qi_p\left(\frac{\Upsilon}{y}\right)
=\Upsilon^{-s}\left(M_s-M_s\left(\frac{\Upsilon}{\gamma_{\min}}\right)\right),
\end{equation*}
The relation
\begin{equation*}
\label{UpsilonOverGammaMin}
\frac{\Upsilon}{\gamma_{\min}}\geqslant\Lambda\gg1
\end{equation*}
allows us to simplify the calculations and to use the asymptotic representation of the Macdonald function at large values of the argument \citep{NikiforovUvarov}
\begin{equation*}
\label{asymptoticExpressionForMcDonaldFunction}
K_{5/3}(x)\sim\sqrt{\frac{\pi}{2x}}\,e^{-x},\qquad x\rightarrow\infty.
\end{equation*}
Using this relation, we obtain an asymptotic expression for the incomplete moments of the Macdonald function and, in particular, for the function $\varphi(x)$:
\begin{equation}
\label{asymptoticExpressionForIncompleteMoments}
M_s(x)\sim\sqrt{\frac{\pi}{2}}\,x^{s-1/2}e^{-x},\qquad x\rightarrow\infty,
\end{equation}
\begin{equation}
\label{asymptoticExpressionForPhi}
\varphi(x)\equiv M_0(x)\sim\sqrt{\frac{\pi}{2x}}\,e^{-x},\qquad x\rightarrow\infty.
\end{equation}
Using Eqs.~\eqref{transformedI}--\eqref{asymptoticExpressionForPhi}, we then obtain
\begin{equation*}
\label{prefinalI}
I\approx\Upsilon^{-s}M_s
\end{equation*}
or, representing $I$ as a linear integral operator $I[qi_p(x)](\gamma_\parallel)$ (see Eq.~\eqref{transformedI}),
\begin{equation}
\label{finalI}
I[x^{-s}](y)\approx L(s)y^{-s},
\end{equation}
where
\begin{equation*}
\label{L}
L(s)=\left(\frac{a^2}{\Lambda}\frac{\cos^2\eta}{\sin\eta}\right)^{-s}M_s.
\end{equation*}

Let there be some function $f(x)$ defined in the interval~$(0,\infty)$. This function can be associated with a function
\begin{equation}
\label{directMellinTransform}
g(s)=\mathfrak{M}[f(x)](s)=
\int\limits_0^\infty x^{s-1}f(x)\,dx,
\end{equation}
where $s=\sigma+i\tau$ is a complex parameter. The integral transform \eqref{directMellinTransform} is called the Mellin transform of the function~$f(x)$. If the transform $g(s)$ is known, then the inverse transform $f(x)$ can be found from the formula for the inverse Mellin transform:
\begin{equation}
\label{inverseMellinTransform}
f(x)=\mathfrak{M}^{-1}[g(s)](x)=
\frac{1}{2\pi i}\int\limits_{\sigma-i\infty}^{\sigma+i\infty} x^{-s}g(s)\,ds,
\end{equation}
where $\sigma=\mathrm{Re}\,s$ and $x>0$. The function $f(x)$ must satisfy some very weak constraints. The following constraints can be taken as the latter \citep{PrudnikovBrychkovMarichev1989}. Let~$f(x)$ be absolutely integrable in any finite interval $(\varepsilon_1,\varepsilon_2)$, $0<\varepsilon_1<\varepsilon_2<\infty$, and satisfy the constraints $|f(x)|<Ax^{-\sigma_1}$ for $0<x\leqslant\varepsilon_1$ and $|f(x)|<Ax^{-\sigma_2}$ for $x\geqslant\varepsilon_2$, $\sigma_1<\sigma_2$, $A=\mathrm{const}>0$. The transform $g(s)$ then exists and is an analytic function in the band $\sigma_1<\mathrm{Re}\,s<\sigma_2$. Formula~\eqref{inverseMellinTransform} is valid at each continuity point of the function~$f(x)$. The integration in Eq.~\eqref{inverseMellinTransform} is over any vertical straight line in the analyticity band and the integral at infinity is understood in the sense of a principal value. Note that there also exist generalizations of the integral transform \eqref{directMellinTransform} to the space of generalized functions \citep{BrychkovPrudnikov1977}.

We can now consider the case of arbitrary functions $q_p(x)$ and $qi_p(x)$ by assuming that
\begin{equation*}
\label{qAndQiAsInverseMellinTransforms}
q_p(x)=\mathfrak{M}^{-1}[\mathcal{M}(s)](x),\qquad
qi_p(x)=\mathfrak{M}^{-1}[\mathcal{M}i(s)](x).
\end{equation*}
The sought-for functions $\mathcal{M}(s)$ and $\mathcal{M}i(s)$ are related by the obvious relation
\begin{equation*}
\label{MandMiRelation}
\mathcal{M}i(s)=\frac{\mathcal{M}(s)}{s}.
\end{equation*}
The integral equation \eqref{integralEquationForFindingPairSource}, once it has been represented as
\begin{equation}
\label{transformedIntegralEquation}
\mathfrak{M}^{-1}[\mathcal{M}(s)](\gamma_\parallel)=\frac{2a}{\Lambda}q_{curv}\left(2\gamma_\parallel\frac{a}{\Lambda}\right)
+\frac{a^3}{\Lambda}\xi\int\limits_0^{\eta_i}d\eta\frac{\cos^2\eta}{\sin\eta}
\,I\left[\mathfrak{M}^{-1}[\mathcal{M}i(s)](x)\right](\gamma_\parallel)
\end{equation}
can be easily solved. The linearity of the operator $I$ and relation \eqref{finalI} allow the following expression to be derived:
\begin{equation}
\label{IwithInverseMellin}
I\left[\mathfrak{M}^{-1}[\mathcal{M}i(s)](x)\right](\gamma_\parallel)
=\mathfrak{M}^{-1}[L(s)\mathcal{M}i(s)](\gamma_\parallel).
\end{equation}
Now, it is necessary to take into account Eq.~\eqref{IwithInverseMellin} and then apply the direct Mellin transform operator $\mathfrak{M}$ to both sides of Eq.~\eqref{transformedIntegralEquation}. Using the formula
\begin{equation}
\label{MellinPhi}
\mathfrak{M}\left[\varphi(x)\right](s)=\frac{M_s}{s}
\end{equation}
and the property
\begin{equation}
\label{MellinProperty}
\mathfrak{M}[f(Ax)](s)=A^{-s}\mathfrak{M}[f(x)](s),\qquad A>0,
\end{equation}
when calculating the Mellin transform of the source of curvature photons $q_{curv}$, we ultimately obtain the expression for the Mellin transform $\mathcal{M}(s)$ of the function $q_p(x)$
\begin{equation}
\label{MellinTransformForQp}
\mathcal{M}(s)=\frac{1}{\sqrt{3}\pi}\frac{\alpha}{\gamma_0^2}\frac{2a}{\Lambda}
\left(\frac{2a}{\Lambda k_{0}}\right)^{-s}\frac{M_s}{s}
\left(1-\frac{a^3}{\Lambda}\xi\left(\frac{a^2}{\Lambda}\right)^{-s}\nu_s\frac{M_s}{s}\right)^{-1}.
\end{equation}
Here, $k_0=3\gamma_0^3/2\rho$ is the characteristic energy of the curvature photons (see~\eqref{curvatureEnergy}) corresponding to $\gamma_\parallel=\gamma_0$ and the function $\nu_s$ is defined by the relation
\begin{equation*}
\label{nuS}
\nu_s=\int\limits_0^{\eta_i} d\eta\left(\frac{\cos^2\eta}{\sin\eta}\right)^{1-s}.
\end{equation*}

It was shown in \citep{SobyaninDissertation} that $\mathcal{M}(s)$ is a meromorphic function and, consequently, has no other singular points, except the poles, in the complex plane. To choose the path of integration in Eq.~\eqref{inverseMellinTransform}, it is necessary to find the analyticity band of the function~$\mathcal{M}(s)$. Obviously, this band must be bounded on the left and the right by the vertical straight lines on which the function has singular points, which are poles in the case under consideration. We will search for the bounding poles on the real axis. In addition, we will be interested in $s>2/3$. The poles will then be specified by the equation
\begin{equation}
\label{equationForPoles}
P(s)=1,
\end{equation}
where we introduced a function
\begin{equation*}
\label{pS}
P(s)=
\frac{a^3}{\Lambda}\xi\left(\frac{a^2}{\Lambda}\right)^{-s}\nu_s\frac{M_s}{s}.
\end{equation*}
Let $s_{\min}$ and $s_{\max}$ be the left and right bounding poles, respectively. The cumbersome calculations performed in~\citep{SobyaninDissertation} allow us to find that
\begin{equation}
\label{sMin}
s_{\min}=2-\frac{\Lambda}{a\ln\Lambda},
\end{equation}
\begin{equation}
\label{sMax}
s_{\max}=2+e\Lambda.
\end{equation}
We see that the left bounding pole is always leftward of $s=2$. The quantity $\Lambda/a\ln\Lambda\approx0.05$. For the right bounding pole, we have $s_{\max}\gg1$. At $\Lambda\sim10$, $s_{\max}\approx30$. The estimated upper limit for the absolute value of the function $P(s)$ leads us to conclude that $|P(s)|<1$ inside the band $s_{\min}<\mathrm{Re}\,s<s_{\max}$ \citep{SobyaninDissertation}. Since relation \eqref{equationForPoles} holds at the poles, the function $\mathcal{M}(s)$ inside the same band has no poles. Thus, the function $\mathcal{M}(s)$ is regular in the band $s_{\min}<\mathrm{Re}\,s<s_{\max}$ and the path of integration in Eq.~\eqref{inverseMellinTransform} can be taken in the form of a vertical straight line parallel to the imaginary axis and located inside this band.

Before we calculate the function $Q(\gamma_{\min})$, let us investigate the properties of the produced-particle spectrum~$q_p(x)$. We can consider the behavior of this
function at low and high values of the argument. We will be interested primarily in the low-energy asymptotics of the spectrum. In order to have the right to consider it, we will assume the existence of some transition Lorentz factor~$\gamma_{tr}$ that separates the entire spectrum into two regions: the region of low energies and the region of high energies. Since the entire spectrum extends from~$\gamma_{\min}$ to~$\infty$, for such a separation to be possible, we must assume that at least $\gamma_{\min}<\gamma_{tr}$.

Let us calculate the low-energy asymptotics of $q_p(x)$. For this purpose, it is convenient to express the Mellin transform $\mathfrak{M}$ in terms of the equivalent two-sided Laplace transform $\mathfrak{L}$ specified by the formula
\begin{equation*}
\label{directLaplaceTransform}
g(s)=\mathfrak{L}[f(x)](s)=\int\limits_{-\infty}^\infty e^{-sx}f(x)\,dx.
\end{equation*}
The corresponding inverse two-sided Laplace transform is
\begin{equation}
\label{inverseTwosideLaplaceTransform}
f(x)=\mathfrak{L}^{-1}[g(s)](x)=\frac{1}{2\pi i}\int\limits_{\sigma-i\infty}^{\sigma+i\infty} e^{xs}g(s)\,ds.
\end{equation}
Then,
\begin{equation*}
\label{qPviaInverseTwosideLaplaceTransform}
q_p(x)=\mathfrak{L}^{-1}[\mathcal{M}(s)](-\ln x).
\end{equation*}
Consider the range of low $x$. If $x\rightarrow0$, then $-\ln x>0$ and, moreover, $-\ln x\gg1$. As follows from Eq.~\eqref{inverseTwosideLaplaceTransform}, it is advantageous to have $s$ such that $\mathrm{Re}(s)$ are as small as possible. This can be achieved by transferring the path of integration as far to the left as possible. Obviously, the integral is invariable in this case if the path is transferred within the regularity band. To obtain the principal term in the asymptotic expansion of the function $q_p(x)$ at~$x=0$, the path should be transferred through the left bounding pole~$s_{\min}$. The original integral will then be equal to the sum of the residue of the integrand at~$s=s_{\min}$ multiplied by~$2\pi i$ and the integral over the transferred path. The former term will give the principal term of the asymptotic expansion, while the latter will have the next order of smallness. We can now write
\begin{equation*}
\label{qPzeroAsymptotics}
q_p(x)\sim\underset{s=s_{\min}}{\mathrm{res}}
x^{-s}\mathcal{M}(s),\qquad x\rightarrow0.
\end{equation*}
Formula \eqref{sMin} allows the relation
\begin{equation*}
\label{sEquals2}
s_{\min}\approx2
\end{equation*}
to be used in the calculations.

Turning to Eq.~\eqref{MellinTransformForQp}, we will obtain the final expression for the low-energy asymptotics of the function~$q_p(x)$:
\begin{equation}
\label{qPlowEnergyAsymptotics}
q_p(x)\sim\frac{\alpha}{\sqrt{3}\pi}\frac{\Lambda}{2a}
\left(\frac{k_0}{\gamma_0}\right)^2\frac{x^{-2}}{\xi\ln\Lambda},\qquad x\rightarrow0.
\end{equation}

Recall, however, that the function $q_p(x)$ is defined at~$x\geqslant\gamma_{\min}\gg1$. Therefore, it is necessary to understand why Eq.~\eqref{qPlowEnergyAsymptotics} can be used in this case. The expression $x\rightarrow0$ implies that the values of the argument under consideration are small compared to~$\gamma_{tr}$. Recall that $|P(s)|<1$ in the regularity band. To qualitatively estimate~$\gamma_{tr}$ we will set $P(s)=0$ in Eq.~\eqref{MellinTransformForQp}, which formally corresponds to~$\xi=0$. It follows from Eqs.~\eqref{MellinPhi} and~\eqref{MellinProperty} that the inverse transform of the derived function is proportional to the function $\varphi(x/\gamma_{tr})$, where
\begin{equation*}
\label{transitionalLorentzFactorInAsymptotics}
\gamma_{tr}=\frac{k_0\Lambda}{2a}.
\end{equation*}
Since the transition region for the function $\varphi(y)$ is $y\approx1$, the region of low energies is specified by the inequalities $\gamma_{\min}<\gamma_\parallel<\gamma_{tr}$. For a proper consideration, this region should not be vanishingly small. Therefore, it should be assumed that
\begin{equation}
\label{asymptoticsCorrectness}
\gamma_{\min}\ll\gamma_{tr}.
\end{equation}
For typical values of $k_0\sim10^5$, $a\sim100$, and $\Lambda\sim10$, we have a characteristic transition Lorentz factor $\gamma_{tr}\sim10^4$. We see that criterion \eqref{asymptoticsCorrectness} definitely holds for a typical minimum Lorentz factor $\gamma_{\min}\sim100$.

As regards the high-energy asymptotics, we will not need it in the subsequent calculations. If necessary, it can be obtained by a method similar to that described above, but the path of integration should be transferred through the right bounding pole. Note only that the high-energy asymptotics is proportional to~$x^{-s_{\max}}$, where $s_{\max}$ is defined by Eq.~\eqref{sMax}. It has a power-law form rather than an exponential one, consistent with the conditions described after Eq.~\eqref{inverseMellinTransform}.

Finally, let us calculate the required function $Q(\gamma_{\min})$ \eqref{normalizedTotalPlasmaSource}. No direct integration of the function $q_p(x)$ is possible. However, the Mellin transform of the function $Q(x)$ can be calculated using the relation
\begin{equation*}
\label{QxMellinTransform}
\mathcal{M}^{int}(s)=\mathfrak{M}[Q(x)](s)=\frac{\mathcal{M}(s+1)}{s}.
\end{equation*}
For the function $Q(x)$, we then have
\begin{equation*}
\label{QxViaMellinTransform}
Q(x)=\mathfrak{M}^{-1}[\mathcal{M}^{int}(s)](x).
\end{equation*}
The function $\mathcal{M}^{int}(s)$ is meromorphic in the entire complex plane and regular in the band $s_{\min}^{int}<\sigma<s_{\max}^{int}$, where $s_{\min}^{int}=s_{\min}-1$ and $s_{\max}^{int}=s_{\max}-1$ are the left and right bounding poles, respectively. It remains to transfer the path of integration beyond the left bounding pole $s_{\min}^{int}\approx1$ and to find that
\begin{equation*}
\label{QxLowEnergyAsymptotics}
Q(x)\sim xq_p(x),\qquad x\rightarrow0.
\end{equation*}
It makes sense to consider this asymptotic expression at~$x\ll\gamma_{tr}$. Using criterion \eqref{asymptoticsCorrectness} and taking $\gamma_{\min}$ as~$x$, we
have
\begin{equation}
\label{QxViaGammaAndRho}
Q(\gamma_{\min})=\frac{\alpha}{3}\frac{\Lambda}{a}\frac{\gamma_0^4}{\rho^2}
\frac{\gamma_{\min}^{-1}}{\ln\Lambda}.
\end{equation}
It is easy to see that the source of electron-positron pairs $Q$ depends linearly on the total intensity of the curvature radiation \eqref{curvatureRadiationIntensity}:
\begin{equation}
\label{QviaCurvatureIntensity}
Q=\lambda W_{curv}B,
\end{equation}
where
\begin{equation*}
\label{coefLambda}
\lambda=\frac{3}{8}\frac{\Lambda}{\ln\Lambda}\frac{1}{\gamma_{\min}}\ll1.
\end{equation*}
In order of magnitude, $\lambda\sim0.1$.

Above, we considered the production of electron-positron pairs by curvature and synchrotron photons. The presence of synchrotron photons is important when considering an electron-positron cascade in the polar cap for a closed magnetosphere completely filled with plasma. In this case, even despite the electric field screening, apart from the primary particles accelerated in the polar gap, there are two generations of secondary particles: the first generation of secondary particles produced by the curvature photons from primary particles and the second generation of particles produced by the synchrotron photons emitted through the transition of first-generation particles to the zeroth Landau level. In our case, we have a significant longitudinal electric field that continuously accelerates the newly produced particles. For this reason, it makes no sense to consider the plasma production cascade as a set of successive generations, because the produced particles, whatever generation they formally belong to, immediately begin to play the role of primary particles after their acceleration. Consequently, the presence of synchrotron photons in the case we consider is all the more important, because we formally have the case corresponding to an infinite number of particle generations. In other words, the second term on the right-hand side of the integral equation~\eqref{integralEquationForFindingPairSource} is much larger than the first one. From qualitative considerations, we can write the inequality
\begin{equation}
\label{synchrotronPhotonsInequality}
Q\gg\frac{W_{curv}}{k_0},
\end{equation}
with the characteristic number of curvature photons produced per unit time appearing on its right-hand side. Inequality~\eqref{synchrotronPhotonsInequality} means that the existing synchrotron radiation makes a major contribution to the total pair production rate. Using Eqs.~\eqref{QviaCurvatureIntensity} and~\eqref{synchrotronPhotonsInequality}, we obtain
\begin{equation}
\label{synchrotronDomination}
\gamma_{\min}\ll\frac{\gamma_{tr}}{\ln\Lambda}.
\end{equation}
In particular, given the inequality $\ln\Lambda>1$, condition~\eqref{asymptoticsCorrectness} follows from inequality~\eqref{synchrotronDomination}. Since~$\Lambda\sim10$, inequalities~\eqref{asymptoticsCorrectness} and~\eqref{synchrotronDomination} may be considered to be essentially equivalent. They both always hold, because it is implied that inequality \eqref{synchrotronPhotonsInequality} holds. Otherwise, in the presence of only curvature radiation, we should have formally set $\xi=0$ in Eq.~\eqref{integralEquationForFindingPairSource}, which would immediately give the solution.

When the condition of quasi-stationary motion $W_{curv}=E_\parallel$ is met, we have
\begin{equation}
\label{quasistationaryQ}
Q=\lambda\,|\mathbf{E}\cdot\mathbf{B}|.
\end{equation}

\section{THE INITIAL STAGE OF CASCADE DEVELOPMENT}

Let there be a primary electron-positron pair produced by a photon from the external cosmic gamma-ray background in a neutron star magnetosphere. In the presence of an external longitudinal electric field, the particles moving virtually along a magnetic field line begin to fly apart in opposite directions and to undergo acceleration, reaching the Lorentz factor~$\gamma_0$. The emerging curvature radiation will be absorbed in the magnetic field through the production of secondary electron-positron pairs. We cannot use Eq.~\eqref{QxViaGammaAndRho} from the initial time~$t=0$, because this requires that the particle number density changes little in the photon mean free path. Consequently, to be able to consider the electron-positron plasma generation processes using the hydrodynamic equations at least requires that a time equal in dimensionless units to the photon mean free path elapses since the initial time~$t=0$. The particles of the electron-positron pair produced by this photon will then give secondary curvature and synchrotron radiation, which also produces pairs and the source of particles from the integral equation~\eqref{integralEquationForFindingPairSource} can subsequently be used.

Let us find the characteristic distance $l_1$ from the production point of a primary electron-positron pair to the production point of the first secondary electron-positron pair. It can be represented as the sum of three parts,
\begin{equation*}
\label{firstSecondaryPairCreationDistance}
l_1=l_a+l_{e}+l_{curv},
\end{equation*}
where $l_a=\gamma_1/E$ is the distance at which the particle acquires some Lorentz factor $\gamma_1\gg\gamma_\parallel$, $l_{e}=\rho/\gamma_1$ is the formation length of curvature radiation, and $l_{curv}= l_f(k_{curv})$ is the mean free path of the curvature photon emitted by the particle. Here, we introduced the mean free path
\begin{equation*}
\label{generalMFP}
l_f(k)=\frac{2\rho}{k}\frac{a}{\Lambda}
\end{equation*}
dependent on the photon energy $k$ \citep{IstominSobyanin2007}. The distance $l_1$ is a function of the Lorentz factor~$\gamma_1$. The minimum value should be chosen from the various $l_1$, because efficient plasma multiplication begins at times $t>\min\limits_{\gamma_1}l_1(\gamma_1)$. The extremality condition $dl_1/d\gamma_1=0$ gives the Lorentz factor
\begin{equation}
\label{gamma1}
\gamma_1=(2\rho)^{1/2}\left(\frac{aE_\parallel}{\Lambda}\right)^{1/4}
\end{equation}
and the distance
\begin{equation}
\label{l1}
l_1=\frac{4}{3}\frac{\gamma_1}{E_\parallel}.
\end{equation}
Formulas \eqref{gamma1} and \eqref{l1} are valid under the following condition:
\begin{equation}
\label{gamma1andL1ApplicabilityCriterion}
\sqrt{\frac{\Lambda E_\parallel}{a}}\ll1.
\end{equation}
Since the expression on the left-hand side of inequality \eqref{gamma1andL1ApplicabilityCriterion} at $E_\parallel\sim10^{-6}$, $B\sim10^{-2}$, and $\Lambda\sim10$ is equal in order of magnitude to $10^{-4}-10^{-3}$, the inequality itself always holds. Let us take $\rho\sim10^{17}$. The Lorentz factor is then $\gamma_1\approx3\times10^7$ and the corresponding distance is $l_1\approx3\times10^{13}$, which corresponds to $l_1\sim10$~m in dimensional units.

Next, we can consider the production of an electron-positron plasma in the hydrodynamic approximation by specifying linear electron and positron densities in the form
\begin{equation}
\label{electronAndPositronDensityInitialConditions}
N^{init}_\pm=\frac{3}{2l_1}\theta(l_1-|z|),
\end{equation}
where $\theta(x)$ is the theta function, as the initial conditions. It is important to note that the electron and positron densities as functions of the longitudinal coordinate $z$ are determined to within distances of the order of the photon mean free path relative to the electron-positron pair production. In the case under consideration, the distance $l_1$ acts as this length and it does not matter at all which specific function $N^{init}_\pm(z)$ is taken on such small scales. The only requirement is that the integration of this function over the longitudinal coordinate gives the total number of electron-positron pairs. In Eq.~\eqref{electronAndPositronDensityInitialConditions}, we took a uniform distribution of particles in longitudinal coordinate.

It was shown in \citep{SobyaninDissertation} that in the interval from the production time of the first secondary pair $\tau_1\approx l_1$ to the time of full particle acceleration $\tau_{st}$, only a few tens of electron-positron pairs are produced at $E_\parallel\sim10^{-6}$ and~$a/\Lambda\sim10$. The corresponding ratio of the times is $\tau_1/\tau_{st}\approx0.9$. After the time $\tau_{st}$, the particle ceases to accelerate and emits
\begin{equation}
\label{Qc}
Q_{curv}=\frac{5}{2\sqrt{3}}\frac{\alpha\gamma_0}{\rho}
\end{equation}
of curvature photons per unit time. This expression is derived by integrating Eq.~\eqref{summaryCurvatureEmissionProbability} over all energies $k$ from~$0$ to~$\infty$. The characteristic mean free path is given by the expression
\begin{equation*}
\label{curvatureMeanFreePath}
l_{curv}=\frac{8}{9}\alpha\tau_{st}\frac{a}{\Lambda}.
\end{equation*}
For $a/\Lambda\sim10$, the ratio of the curvature photon mean free path to the particle acceleration time is $l_{curv}/\tau_{st}\approx0.06\ll1$. Consequently, once the stationary Lorentz factor $\gamma_0$ has been reached, the production of electron-positron pairs with the source $Q_{curv}$ begins almost immediately. In this case, the ratio of the mean free path to the formation length of the curvature radiation is large, $\sim10^3$; therefore, the radiation formation length may be neglected. $Q_{curv}$~can be taken as the source of electron-positron pairs only at times $\tau_{st}<t<2\tau_1$, because the contribution from the curvature photons produced by secondary particles will subsequently appear.

Thus, the cascade process of electron-positron plasma multiplication will subsequently begin. In this case, the effective switch-on of curvature radiation can be assumed to occur $\tau_{st}$ after the particle production if $\gamma_\parallel\ll\gamma_0$. It is important to note that the minimum mean free path for synchrotron photons at $\Lambda\sim10$ coincides (in dimensionless units) in order of magnitude with the time of full particle acceleration \citep{SobyaninDissertation}. Hence, the production of secondary electron-positron pairs by the curvature and synchrotron photons emitted by the particle begins almost simultaneously.

\section{THE TIME DELAY AND THE EFFECTIVE LOCAL SOURCE}

It can be concluded from the preceding section that the particles of each newly produced electron-positron pair contribute to the total pair production rate some delay time $\tau$ after its production. We can then write the differential-difference equation
\begin{equation}
\label{equationForPairGenerationWithTimeDelay}
\frac{dN_\Sigma(t)}{dt}=2QN_\Sigma(t-\tau)
\end{equation}
with the initial conditions
\begin{equation*}
\label{initialConditionsForDifferenceDifferentialEquation}
N_\Sigma(t)=N_0,\qquad 0\leqslant t\leqslant \tau,
\end{equation*}
where $N_0$ is the number of electron-positron pairs at the initial time. Here, we denote the total number of electron-positron pairs by
\begin{equation*}
\label{notationForIntegratedDensities}
N_\Sigma(t)=\int\limits_{-\infty}^\infty N_\pm(z,t)dz,
\end{equation*}
where $N_+(z,t)$ and $N_-(z,t)$ are the linear positron and electron densities, respectively. Let~$N_\tau= 2Q\tau$ be the number of particles produced in the time $\tau$ by the photons emitted by one particle. It is also convenient to pass to the time $t_n= t/\tau$ normalized to the delay time. Equation~\eqref{equationForPairGenerationWithTimeDelay} can be solved using the Laplace transform (see, e.g., \citep{Pinney1961}):
\begin{equation}
\label{PinneySolution}
N_\Sigma(t)=N_0\sum\limits_{p=0}^{[t_n]}\frac{N_\tau^p}{p!}(t_n-p)^p,
\qquad t\geqslant \tau.
\end{equation}
This exact solution allows the total number of electron-positron pairs to be found but does not allow the distribution of electrons and positrons in longitudinal coordinate to be found. In addition, it is applicable until the screening of the external longitudinal electric field begins. The hydrodynamic equations suggest the existence of a dependence of the pair production rate at some point $z$ and at some instant of time $t$ on the electron and positron densities taken at the same point and at the same instant of time. In other words, the equations under consideration are local in space and time. Nevertheless, there exist both spatial and temporal separations of the photon production and absorption points. If the change in electron and positron densities on characteristic photon mean free paths is small, then the spatial nonlocality condition imposes no significant constraints on the applicability of the hydrodynamic equations, except the impossibility of considering the plasma generation on short time scales (see the preceding section). In contrast, the temporal nonlocality condition is significant, because, as we will see below, a large number of photons are produced in the delay time:
\begin{equation*}
\label{Qtau}
N_\tau\gg1.
\end{equation*}
Consequently, disregarding the temporal nonlocality would cause the electron-positron pair production rate to be grossly overestimated.

We need to find some effective local source $Q^{eff}$ that can be used in the hydrodynamic equations. To calculate~$Q^{eff}$, we will assume that all of the parameters characterizing the plasma tube change little at a distance of the order of the photon mean free path~$l_f$. We can then select some tube segment with a length equal to several $l_f$ in which the local homogeneity condition is met. For the tube segment under consideration, the source $Q$ and the linear electron and positron densities $N_\pm$ are
essentially invariable as functions of the longitudinal coordinate~$z$. For the total number of electron-positron pairs in the tube segment under consideration, we can then write Eq.~\eqref{equationForPairGenerationWithTimeDelay}.

Let us investigate solution \eqref{PinneySolution}. Let us introduce a function
\begin{equation*}
\label{SonP}
S(p)= p\ln(e N_\tau \xi(p)),
\end{equation*}
where
\begin{equation*}
\label{ZonP}
\xi(p)=\frac{t_n}{p}-1.
\end{equation*}
Replacing the summation by the integration over the
variable~$p$ in Eq.~\eqref{PinneySolution} leads to an asymptotic estimate of the series
\begin{equation}
\label{asymptoticEstimateForPinneySeries}
N^{ap}_\Sigma(t)\sim N_0\int\limits_0^{t_n}\frac{1}{\sqrt{2\pi p}}\exp\bigl(S(p)\bigr)\,dp.
\end{equation}
Let $p_0$ be the point of maximum of the function~$S(p)$, i.e.,
\[
S(p_0)=\max\limits_{0\leqslant p\leqslant t_n}S(p).
\]
The equation
\begin{equation}
\label{eqOnXiAtMaximumP}
\frac{1}{\xi_0}\exp\frac{1}{\xi_0}=N_\tau,
\end{equation}
where $\xi_0=\xi(p_0)$, then follows from the extremality condition~$S'(p_0)=0$. At~$N_\tau\gg1$, the inequality $\xi_0\ll1$ holds and $\xi_0$ can be easily found by the method of successive approximations:
\begin{equation*}
\label{iterativeSolutionXi}
\xi_0\approx[\ln N_\tau-\ln(\ln N_\tau)]^{-1}.
\end{equation*}
We then have the point of maximum
\begin{equation*}
\label{maxP0}
p_0=\frac{t_n}{1+\xi_0}.
\end{equation*}
We can now estimate the integral in Eq.~\eqref{asymptoticEstimateForPinneySeries} by the
Laplace method \citep{Fedoryuk1977} to obtain
\begin{equation}
\label{finalNestimateFromDDeq}
N^{ap}_\Sigma(t)=\frac{N_0\xi_0}{1+\xi_0}\bigl(e N_\tau \xi_0\bigr)^{p_0}.
\end{equation}
The figure shows the time dependence of $\bigl(N^{ap}_\Sigma(t)-N_\Sigma(t)\bigr)/N_\Sigma(t)$ for~$N_\tau=10^5$. We see that our asymptotic estimate under the condition $t_n\gg1$ closely approaches the exact solution, with the error tending to zero as $t$ increases. At lower~$N_\tau$, the relative error only decreases. The function $N_\Sigma(t)$ is only continuous rather than infinitely differentiable. This is due to the stepwise ``switch-on'' of the next generation of particles contributing to the total electron-positron pair production rate at time intervals~$\tau$. The most important thing here is that the plot of $N^{ap}_\Sigma(t)$ intersects the plot of $N_\Sigma(t)$ in an arbitrarily chosen interval $n\tau<t<(n+1)\tau$, $n=0$, $1$, $2$, \ldots. Consequently, the derived smooth monotonic approximation $N^{ap}_\Sigma(t)$ smoothes~$N_\Sigma(t)$. Even differing from the exact solution in certain time intervals, it properly describes the mean rate of increase in the number of particles and allows the error accumulation in the number of produced pairs on long time scales to be avoided.

Differentiating Eq.~\eqref{finalNestimateFromDDeq} with respect to time using relation \eqref{eqOnXiAtMaximumP} gives
\begin{equation*}
\label{finaldNdt}
\frac{dN^{ap}_\Sigma(t)}{dt}=2Q^{eff}N^{ap}_\Sigma(t),
\end{equation*}
where we specified the effective source of electron-positron pairs
\begin{equation}
\label{effectivePairSource}
Q^{eff}=\frac{1}{2\xi_0\tau}.
\end{equation}
Thus, the number of electron-positron pairs will increase exponentially with the effective local source $Q^{eff}$ in each selected segment of the plasma tube.

It remains to determine the delay time $\tau$ and the initial source~$Q$. Consider one of the particles of the produced electron-positron pair. If we exclude the synchrotron photons from consideration, then the curvature photons emitted by the particle will begin to contribute to the production of secondary electron-positron pairs after the time $\tau_{st}$ of full particle acceleration. In this case, $Q_{curv}$~defined by Eq.~\eqref{Qc} should be taken as the source~$Q$. This case corresponds to the presence of a strong magnetic field $B\gtrsim1$. If, alternatively, we also include the synchrotron photons in consideration, then, as follows from the preceding section, they will begin to contribute to the pair production on time scales of the same order of magnitude. In addition, when deriving the integral equation~\eqref{integralEquationForFindingPairSource}, we assumed that each newly produced particle accelerated to the Lorentz factor $\gamma_0$ and began to act as the source of primary curvature photons in the expression for~$q_{curv}$. This will also occur the time $\tau_{st}$ after the particle production. Before this instant of time, the particle under consideration is not involved in the production of secondary electron-positron pairs and, consequently, makes no contribution in the form of the fraction of the total pair production rate per particle defined by~\eqref{QxViaGammaAndRho} and Eq.~\eqref{quasistationaryQ} following from it. Thus, Eq.~\eqref{quasistationaryQ} and the time $\tau_{st}$ should be taken as the source $Q$ and the delay time~$\tau$, respectively. This case corresponds to the presence of a weak magnetic field $B\ll1$. We then have
\begin{equation*}
\label{Ntau0}
N_{\tau_{st}}=2\lambda\gamma_0 B\gg1.
\end{equation*}
At $B\sim0.01$ and $\gamma_0\sim10^8$, we obtain $N_{\tau_{st}}\sim10^5$, i.e., the inequality $N_{\tau_{st}}\gg1$ holds. The corresponding value of $\xi_0\approx0.1$. Note that when the electric field is subsequently screened, this inequality will also hold even if the longitudinal Lorentz factor gradually decreases down to the small value $\gamma_0\sim10^3$. The required effective local source will be defined by Eq.~\eqref{effectivePairSource}, in which $\tau=\tau_{st}$ and $\xi_0$ is the solution of Eq.~\eqref{eqOnXiAtMaximumP} for~$N_\tau=N_{\tau_{st}}$. Finally, we can write
\begin{equation}
\label{finalQeff}
Q^{eff}=\frac{\alpha}{3\,\xi_0}\frac{\gamma_0^3}{\rho^2}.
\end{equation}

\section{CONCLUSIONS}

The absorption of a high-energy photon from the external cosmic gamma-ray background in the inner neutron star magnetosphere triggers the generation of a secondary electron-positron plasma. The existence of a strong longitudinal electric field is a necessary condition for the development of a cascade. Allowance for the pair production by curvature and synchrotron photons in a weak (compared to the Schwinger one) magnetic field necessitates considering the integral equation~\eqref{integralEquationForFindingPairSource}. The contribution from synchrotron photons to the source of electron-positron pairs in the case of a weak magnetic field is decisive. In this case, the contribution from bremsstrahlung photons may be neglected. The integral equation is solved using the Mellin transform. The transfer of the path of integration beyond the pole bounding the analyticity band of the transform of the solution on the left and the method of residues allow one to obtain the low-energy asymptotics of the solution and to show its applicability for typical conditions in the neutron star magnetosphere. If the conditions of quasi-stationary motion are met, then the solution is reduced to Eq.~\eqref{quasistationaryQ}. In the case of a strong magnetic field $B\gtrsim1$, because of the absence of synchrotron photons, the source of electron-positron pairs is determined by the intensity of the curvature radiation and has the form~\eqref{Qc}. Immediately after the absorption of the primary gamma-ray photon triggering a cascade in the magnetosphere, the hydrodynamic equations are inapplicable. The particles of the produced primary electron-positron pair begin to move in opposite directions almost along a magnetic field line and reach a Lorentz factor of $\sim10^8$ in the initially vacuum electric field. The charged particle acceleration time, the curvature radiation formation length, and the photon mean free path relative to the electron-positron pair production are finite. This determines the production of secondary pairs at a distance of $\sim10$~m from the primary pair. This distance is comparable to the distance of particle acceleration to a stationary Lorentz factor. Subsequently, the next generation stage begins, when the pair production occurs due to intense absorption of the curvature photons produced by the particles of the primary pair during their acceleration. Since the production rate of curvature photons by a fully accelerated particle is high, the effective switch-on of curvature radiation occurs $\tau_{st}$ after the electron-positron pair production. The same is also true of the synchrotron photons: their minimum mean free path is comparable to the distance at which the particle reaches a stationary Lorentz factor. The absorption of the synchrotron photons produced by the particles of secondary electron-positron pairs when passing to the zeroth Landau level gives rise to the main stage of electron-positron plasma generation when its rate is highest. The high secondary plasma generation rate due to the existence of a strong accelerating electric field necessitates allowance for the time delay~$\tau$. After this time, the particles of each newly produced electron-positron pair begin to contribute to the total pair production rate through the absorption of the curvature and synchrotron photons emitted by them in the magnetosphere. Solving the differential-difference equation \eqref{equationForPairGenerationWithTimeDelay} allows one to calculate the effective local source of electron-positron pairs~\eqref{finalQeff}, using which the time delay can be properly taken into account. Significantly, the delay leads to a great reduction in the electron-positron plasma generation rate compared to the case of zero time delay. The effective local source \eqref{finalQeff} is much weaker than both the initial source \eqref{Qc} for $B\gtrsim1$, when there is no synchrotron radiation, and the initial source \eqref{quasistationaryQ} for $B\ll1$, when the intensity of the synchrotron radiation exceeds that of the curvature one. In this case, the delay time $\tau$ for both weak and strong magnetic fields is almost the same and is comparable in order of magnitude to the particle transition time $\tau_{st}$ to a quasi-stationary regime of motion.

A detailed study of the formation of a plasma tube in a neutron star magnetosphere through the cascade multiplication of an electron-positron plasma triggered by the absorption of a high-energy photon from the external cosmic gamma-ray background with allowance made for the screening of the external electric field requires a separate consideration.

\begin{acknowledgments}
This work was supported in part by the Russian Foundation for Basic Research (project no.~11-02-01021-a).
\end{acknowledgments}
\newpage
\bibliography{JETP1}

\providecommand{\noopsort}[1]{}\providecommand{\singleletter}[1]{#1}%
\begin{thebibliography}{35}%
\makeatletter
\providecommand \@ifxundefined [1]{%
 \@ifx{#1\undefined}
}%
\providecommand \@ifnum [1]{%
 \ifnum #1\expandafter \@firstoftwo
 \else \expandafter \@secondoftwo
 \fi
}%
\providecommand \@ifx [1]{%
 \ifx #1\expandafter \@firstoftwo
 \else \expandafter \@secondoftwo
 \fi
}%
\providecommand \natexlab [1]{#1}%
\providecommand \enquote  [1]{``#1''}%
\providecommand \bibnamefont  [1]{#1}%
\providecommand \bibfnamefont [1]{#1}%
\providecommand \citenamefont [1]{#1}%
\providecommand \href@noop [0]{\@secondoftwo}%
\providecommand \href [0]{\begingroup \@sanitize@url \@href}%
\providecommand \@href[1]{\@@startlink{#1}\@@href}%
\providecommand \@@href[1]{\endgroup#1\@@endlink}%
\providecommand \@sanitize@url [0]{\catcode `\\12\catcode `\$12\catcode
  `\&12\catcode `\#12\catcode `\^12\catcode `\_12\catcode `\%12\relax}%
\providecommand \@@startlink[1]{}%
\providecommand \@@endlink[0]{}%
\providecommand \url  [0]{\begingroup\@sanitize@url \@url }%
\providecommand \@url [1]{\endgroup\@href {#1}{\urlprefix }}%
\providecommand \urlprefix  [0]{URL }%
\providecommand \Eprint [0]{\href }%
\providecommand \doibase [0]{http://dx.doi.org/}%
\providecommand \selectlanguage [0]{\@gobble}%
\providecommand \bibinfo  [0]{\@secondoftwo}%
\providecommand \bibfield  [0]{\@secondoftwo}%
\providecommand \translation [1]{[#1]}%
\providecommand \BibitemOpen [0]{}%
\providecommand \bibitemStop [0]{}%
\providecommand \bibitemNoStop [0]{.\EOS\space}%
\providecommand \EOS [0]{\spacefactor3000\relax}%
\providecommand \BibitemShut  [1]{\csname bibitem#1\endcsname}%
\let\auto@bib@innerbib\@empty
%</preamble>
\bibitem [{\citenamefont {Kramer}\ \emph {et~al.}(2006)\citenamefont {Kramer},
  \citenamefont {Lyne}, \citenamefont {O'Brien}, \citenamefont {Jordan},\ and\
  \citenamefont {Lorimer}}]{KramerEtal2006}%
  \BibitemOpen
  \bibfield  {author} {\bibinfo {author} {\bibfnamefont {M.}~\bibnamefont
  {Kramer}}, \bibinfo {author} {\bibfnamefont {A.~G.}\ \bibnamefont {Lyne}},
  \bibinfo {author} {\bibfnamefont {J.~T.}\ \bibnamefont {O'Brien}}, \bibinfo
  {author} {\bibfnamefont {C.~A.}\ \bibnamefont {Jordan}}, \ and\ \bibinfo
  {author} {\bibfnamefont {D.~R.}\ \bibnamefont {Lorimer}},\ }\href@noop {}
  {\bibfield  {journal} {\bibinfo  {journal} {Science (Washington)}\ }\textbf
  {\bibinfo {volume} {312}},\ \bibinfo {pages} {549} (\bibinfo {year}
  {2006})}\BibitemShut {NoStop}%
\bibitem [{\citenamefont {McLaughlin}\ \emph {et~al.}(2006)\citenamefont
  {McLaughlin}, \citenamefont {Lyne}, \citenamefont {Lorimer}, \citenamefont
  {Kramer}, \citenamefont {Faulkner}, \citenamefont {Manchester}, \citenamefont
  {Cordes}, \citenamefont {Camilo}, \citenamefont {Possenti}, \citenamefont
  {Stairs}, \citenamefont {Hobbs}, \citenamefont {D'Amico}, \citenamefont
  {Burgay},\ and\ \citenamefont {O'Brien}}]{McLaughlinEtal2006}%
  \BibitemOpen
  \bibfield  {author} {\bibinfo {author} {\bibfnamefont {M.~A.}\ \bibnamefont
  {McLaughlin}}, \bibinfo {author} {\bibfnamefont {A.~G.}\ \bibnamefont
  {Lyne}}, \bibinfo {author} {\bibfnamefont {D.~R.}\ \bibnamefont {Lorimer}},
  \bibinfo {author} {\bibfnamefont {M.}~\bibnamefont {Kramer}}, \bibinfo
  {author} {\bibfnamefont {A.~J.}\ \bibnamefont {Faulkner}}, \bibinfo {author}
  {\bibfnamefont {R.~N.}\ \bibnamefont {Manchester}}, \bibinfo {author}
  {\bibfnamefont {J.~M.}\ \bibnamefont {Cordes}}, \bibinfo {author}
  {\bibfnamefont {F.}~\bibnamefont {Camilo}}, \bibinfo {author} {\bibfnamefont
  {A.}~\bibnamefont {Possenti}}, \bibinfo {author} {\bibfnamefont {I.~H.}\
  \bibnamefont {Stairs}}, \bibinfo {author} {\bibfnamefont {G.}~\bibnamefont
  {Hobbs}}, \bibinfo {author} {\bibfnamefont {N.}~\bibnamefont {D'Amico}},
  \bibinfo {author} {\bibfnamefont {M.}~\bibnamefont {Burgay}}, \ and\ \bibinfo
  {author} {\bibfnamefont {J.~T.}\ \bibnamefont {O'Brien}},\ }\href@noop {}
  {\bibfield  {journal} {\bibinfo  {journal} {Nature (London)}\ }\textbf
  {\bibinfo {volume} {439}},\ \bibinfo {pages} {817} (\bibinfo {year}
  {2006})}\BibitemShut {NoStop}%
\bibitem [{\citenamefont {Keane}\ \emph {et~al.}(2010)\citenamefont {Keane},
  \citenamefont {Ludovici}, \citenamefont {Eatough}, \citenamefont {Kramer},
  \citenamefont {Lyne}, \citenamefont {McLaughlin},\ and\ \citenamefont
  {Stappers}}]{KeaneEtal2010}%
  \BibitemOpen
  \bibfield  {author} {\bibinfo {author} {\bibfnamefont {E.~F.}\ \bibnamefont
  {Keane}}, \bibinfo {author} {\bibfnamefont {D.~A.}\ \bibnamefont {Ludovici}},
  \bibinfo {author} {\bibfnamefont {R.~P.}\ \bibnamefont {Eatough}}, \bibinfo
  {author} {\bibfnamefont {M.}~\bibnamefont {Kramer}}, \bibinfo {author}
  {\bibfnamefont {A.~G.}\ \bibnamefont {Lyne}}, \bibinfo {author}
  {\bibfnamefont {M.~A.}\ \bibnamefont {McLaughlin}}, \ and\ \bibinfo {author}
  {\bibfnamefont {B.~W.}\ \bibnamefont {Stappers}},\ }\href@noop {} {\bibfield
  {journal} {\bibinfo  {journal} {Mon. Not. R. Astron. Soc.}\ }\textbf
  {\bibinfo {volume} {401}},\ \bibinfo {pages} {1057} (\bibinfo {year}
  {2010})}\BibitemShut {NoStop}%
\bibitem [{\citenamefont {\relax{Yu}. P.~Shitov}\ \emph
  {et~al.}(2009)\citenamefont {\relax{Yu}. P.~Shitov}, \citenamefont {Kuz'min},
  \citenamefont {Dumskii},\ and\ \citenamefont {Losovsky}}]{Shitov2009}%
  \BibitemOpen
  \bibfield  {author} {\bibinfo {author} {\bibnamefont {\relax{Yu}.
  P.~Shitov}}, \bibinfo {author} {\bibfnamefont {A.~D.}\ \bibnamefont
  {Kuz'min}}, \bibinfo {author} {\bibfnamefont {D.~V.}\ \bibnamefont
  {Dumskii}}, \ and\ \bibinfo {author} {\bibfnamefont {B.~Y.}\ \bibnamefont
  {Losovsky}},\ }\href@noop {} {\bibfield  {journal} {\bibinfo  {journal}
  {Astron. Rep.}\ }\textbf {\bibinfo {volume} {53}},\ \bibinfo {pages} {561}
  (\bibinfo {year} {2009})}\BibitemShut {NoStop}%
\bibitem [{\citenamefont {Wang}\ \emph {et~al.}(2007)\citenamefont {Wang},
  \citenamefont {Manchester},\ and\ \citenamefont {Johnston}}]{WangEtal2007}%
  \BibitemOpen
  \bibfield  {author} {\bibinfo {author} {\bibfnamefont {N.}~\bibnamefont
  {Wang}}, \bibinfo {author} {\bibfnamefont {R.~N.}\ \bibnamefont
  {Manchester}}, \ and\ \bibinfo {author} {\bibfnamefont {S.}~\bibnamefont
  {Johnston}},\ }\href@noop {} {\bibfield  {journal} {\bibinfo  {journal} {Mon.
  Not. R. Astron. Soc.}\ }\textbf {\bibinfo {volume} {377}},\ \bibinfo {pages}
  {1383} (\bibinfo {year} {2007})}\BibitemShut {NoStop}%
\bibitem [{\citenamefont {Lyne}\ \emph {et~al.}(2010)\citenamefont {Lyne},
  \citenamefont {Hobbs}, \citenamefont {Kramer}, \citenamefont {Stairs},\ and\
  \citenamefont {Stappers}}]{LyneEtal2010}%
  \BibitemOpen
  \bibfield  {author} {\bibinfo {author} {\bibfnamefont {A.}~\bibnamefont
  {Lyne}}, \bibinfo {author} {\bibfnamefont {G.}~\bibnamefont {Hobbs}},
  \bibinfo {author} {\bibfnamefont {M.}~\bibnamefont {Kramer}}, \bibinfo
  {author} {\bibfnamefont {I.}~\bibnamefont {Stairs}}, \ and\ \bibinfo {author}
  {\bibfnamefont {B.}~\bibnamefont {Stappers}},\ }\href@noop {} {\bibfield
  {journal} {\bibinfo  {journal} {Science (Washington)}\ }\textbf {\bibinfo
  {volume} {329}},\ \bibinfo {pages} {408} (\bibinfo {year}
  {2010})}\BibitemShut {NoStop}%
\bibitem [{\citenamefont {\relax{Yu}. P.~Shitov}\ \emph
  {et~al.}(2000)\citenamefont {\relax{Yu}. P.~Shitov}, \citenamefont
  {Pugachev},\ and\ \citenamefont {Kutuzov}}]{ShitovEtal2000}%
  \BibitemOpen
  \bibfield  {author} {\bibinfo {author} {\bibnamefont {\relax{Yu}.
  P.~Shitov}}, \bibinfo {author} {\bibfnamefont {V.~D.}\ \bibnamefont
  {Pugachev}}, \ and\ \bibinfo {author} {\bibfnamefont {S.~M.}\ \bibnamefont
  {Kutuzov}},\ }\href@noop {} {\bibfield  {journal} {\bibinfo  {journal}
  {Astron. Soc. Pac. Conf. Ser.}\ }\textbf {\bibinfo {volume} {202}},\ \bibinfo
  {pages} {685} (\bibinfo {year} {2000})}\BibitemShut {NoStop}%
\bibitem [{\citenamefont {Malofeev}\ \emph {et~al.}(2005)\citenamefont
  {Malofeev}, \citenamefont {Malov}, \citenamefont {Teplykh}, \citenamefont
  {Tyul'bashev},\ and\ \citenamefont {Tyul'basheva}}]{MalofeevEtal2005}%
  \BibitemOpen
  \bibfield  {author} {\bibinfo {author} {\bibfnamefont {V.~M.}\ \bibnamefont
  {Malofeev}}, \bibinfo {author} {\bibfnamefont {O.~I.}\ \bibnamefont {Malov}},
  \bibinfo {author} {\bibfnamefont {D.~A.}\ \bibnamefont {Teplykh}}, \bibinfo
  {author} {\bibfnamefont {S.~A.}\ \bibnamefont {Tyul'bashev}}, \ and\ \bibinfo
  {author} {\bibfnamefont {G.~E.}\ \bibnamefont {Tyul'basheva}},\ }\href@noop
  {} {\bibfield  {journal} {\bibinfo  {journal} {Astron. Rep.}\ }\textbf
  {\bibinfo {volume} {49}},\ \bibinfo {pages} {242} (\bibinfo {year}
  {2005})}\BibitemShut {NoStop}%
\bibitem [{\citenamefont {Malofeev}\ \emph {et~al.}(2007)\citenamefont
  {Malofeev}, \citenamefont {Malov},\ and\ \citenamefont
  {Teplykh}}]{MalofeevEtal2007}%
  \BibitemOpen
  \bibfield  {author} {\bibinfo {author} {\bibfnamefont {V.~M.}\ \bibnamefont
  {Malofeev}}, \bibinfo {author} {\bibfnamefont {O.~I.}\ \bibnamefont {Malov}},
  \ and\ \bibinfo {author} {\bibfnamefont {D.~A.}\ \bibnamefont {Teplykh}},\
  }\href@noop {} {\bibfield  {journal} {\bibinfo  {journal} {Astrophys. Space
  Sci.}\ }\textbf {\bibinfo {volume} {308}},\ \bibinfo {pages} {211} (\bibinfo
  {year} {2007})}\BibitemShut {NoStop}%
\bibitem [{\citenamefont {Malofeev}\ \emph {et~al.}(2010)\citenamefont
  {Malofeev}, \citenamefont {Teplykh},\ and\ \citenamefont
  {Malov}}]{MalofeevEtal2010}%
  \BibitemOpen
  \bibfield  {author} {\bibinfo {author} {\bibfnamefont {V.~M.}\ \bibnamefont
  {Malofeev}}, \bibinfo {author} {\bibfnamefont {D.~A.}\ \bibnamefont
  {Teplykh}}, \ and\ \bibinfo {author} {\bibfnamefont {O.~I.}\ \bibnamefont
  {Malov}},\ }\href@noop {} {\bibfield  {journal} {\bibinfo  {journal} {Astron.
  Rep.}\ }\textbf {\bibinfo {volume} {54}},\ \bibinfo {pages} {995} (\bibinfo
  {year} {2010})}\BibitemShut {NoStop}%
\bibitem [{\citenamefont {Camilo}\ \emph {et~al.}(2006)\citenamefont {Camilo},
  \citenamefont {Ransom}, \citenamefont {Halpern}, \citenamefont {Reynolds},
  \citenamefont {Helfand}, \citenamefont {Zimmerman},\ and\ \citenamefont
  {Sarkissian}}]{CamiloEtal2006}%
  \BibitemOpen
  \bibfield  {author} {\bibinfo {author} {\bibfnamefont {F.}~\bibnamefont
  {Camilo}}, \bibinfo {author} {\bibfnamefont {S.~M.}\ \bibnamefont {Ransom}},
  \bibinfo {author} {\bibfnamefont {J.~P.}\ \bibnamefont {Halpern}}, \bibinfo
  {author} {\bibfnamefont {J.}~\bibnamefont {Reynolds}}, \bibinfo {author}
  {\bibfnamefont {D.~J.}\ \bibnamefont {Helfand}}, \bibinfo {author}
  {\bibfnamefont {N.}~\bibnamefont {Zimmerman}}, \ and\ \bibinfo {author}
  {\bibfnamefont {J.}~\bibnamefont {Sarkissian}},\ }\href@noop {} {\bibfield
  {journal} {\bibinfo  {journal} {Nature (London)}\ }\textbf {\bibinfo {volume}
  {442}},\ \bibinfo {pages} {892} (\bibinfo {year} {2006})}\BibitemShut
  {NoStop}%
\bibitem [{\citenamefont {Camilo}\ \emph {et~al.}(2007)\citenamefont {Camilo},
  \citenamefont {Ransom}, \citenamefont {Halpern},\ and\ \citenamefont
  {Reynolds}}]{CamiloEtal2007}%
  \BibitemOpen
  \bibfield  {author} {\bibinfo {author} {\bibfnamefont {F.}~\bibnamefont
  {Camilo}}, \bibinfo {author} {\bibfnamefont {S.~M.}\ \bibnamefont {Ransom}},
  \bibinfo {author} {\bibfnamefont {J.~P.}\ \bibnamefont {Halpern}}, \ and\
  \bibinfo {author} {\bibfnamefont {J.}~\bibnamefont {Reynolds}},\ }\href@noop
  {} {\bibfield  {journal} {\bibinfo  {journal} {Astrophys. J. Lett.}\ }\textbf
  {\bibinfo {volume} {666}},\ \bibinfo {pages} {L93} (\bibinfo {year}
  {2007})}\BibitemShut {NoStop}%
\bibitem [{\citenamefont {Levin}\ \emph {et~al.}(2010)\citenamefont {Levin},
  \citenamefont {Bailes}, \citenamefont {Bates}, \citenamefont {Bhat},
  \citenamefont {Burgay}, \citenamefont {Burke-Spolaor}, \citenamefont
  {D'Amico}, \citenamefont {Johnston}, \citenamefont {Keith}, \citenamefont
  {Kramer}, \citenamefont {Milia}, \citenamefont {Possenti}, \citenamefont
  {Rea}, \citenamefont {Stappers},\ and\ \citenamefont {van
  Straten}}]{LevinEtal2010}%
  \BibitemOpen
  \bibfield  {author} {\bibinfo {author} {\bibfnamefont {L.}~\bibnamefont
  {Levin}}, \bibinfo {author} {\bibfnamefont {M.}~\bibnamefont {Bailes}},
  \bibinfo {author} {\bibfnamefont {S.}~\bibnamefont {Bates}}, \bibinfo
  {author} {\bibfnamefont {N.}~\bibnamefont {Bhat}}, \bibinfo {author}
  {\bibfnamefont {M.}~\bibnamefont {Burgay}}, \bibinfo {author} {\bibfnamefont
  {S.}~\bibnamefont {Burke-Spolaor}}, \bibinfo {author} {\bibfnamefont
  {N.}~\bibnamefont {D'Amico}}, \bibinfo {author} {\bibfnamefont
  {S.}~\bibnamefont {Johnston}}, \bibinfo {author} {\bibfnamefont
  {M.}~\bibnamefont {Keith}}, \bibinfo {author} {\bibfnamefont
  {M.}~\bibnamefont {Kramer}}, \bibinfo {author} {\bibfnamefont
  {S.}~\bibnamefont {Milia}}, \bibinfo {author} {\bibfnamefont
  {A.}~\bibnamefont {Possenti}}, \bibinfo {author} {\bibfnamefont
  {N.}~\bibnamefont {Rea}}, \bibinfo {author} {\bibfnamefont {B.}~\bibnamefont
  {Stappers}}, \ and\ \bibinfo {author} {\bibfnamefont {W.}~\bibnamefont {van
  Straten}},\ }\href@noop {} {\bibfield  {journal} {\bibinfo  {journal}
  {Astrophys. J. Lett.}\ }\textbf {\bibinfo {volume} {721}},\ \bibinfo {pages}
  {L33} (\bibinfo {year} {2010})}\BibitemShut {NoStop}%
\bibitem [{\citenamefont {Beskin}(1999)}]{Beskin1999}%
  \BibitemOpen
  \bibfield  {author} {\bibinfo {author} {\bibfnamefont {V.~S.}\ \bibnamefont
  {Beskin}},\ }\href@noop {} {\bibfield  {journal} {\bibinfo  {journal}
  {Phys.---Usp.}\ }\textbf {\bibinfo {volume} {42}},\ \bibinfo {pages} {1071}
  (\bibinfo {year} {1999})}\BibitemShut {NoStop}%
\bibitem [{\citenamefont {Gurevich}\ and\ \citenamefont
  {Istomin}(2007)}]{GurevichIstomin2007}%
  \BibitemOpen
  \bibfield  {author} {\bibinfo {author} {\bibfnamefont {A.~V.}\ \bibnamefont
  {Gurevich}}\ and\ \bibinfo {author} {\bibfnamefont {{\relax Ya}.~N.}\
  \bibnamefont {Istomin}},\ }\href@noop {} {\bibfield  {journal} {\bibinfo
  {journal} {Mon. Not. R. Astron. Soc.}\ }\textbf {\bibinfo {volume} {377}},\
  \bibinfo {pages} {1663} (\bibinfo {year} {2007})}\BibitemShut {NoStop}%
\bibitem [{\citenamefont {Istomin}\ and\ \citenamefont
  {Sobyanin}(2009)}]{IstominSobyanin2009}%
  \BibitemOpen
  \bibfield  {author} {\bibinfo {author} {\bibfnamefont {{\relax Ya}.~N.}\
  \bibnamefont {Istomin}}\ and\ \bibinfo {author} {\bibfnamefont {D.~N.}\
  \bibnamefont {Sobyanin}},\ }\href@noop {} {\bibfield  {journal} {\bibinfo
  {journal} {JETP}\ }\textbf {\bibinfo {volume} {109}},\ \bibinfo {pages} {393}
  (\bibinfo {year} {2009})}\BibitemShut {NoStop}%
\bibitem [{\citenamefont {Istomin}\ and\ \citenamefont
  {Sob'yanin}(2010{\natexlab{a}})}]{IstominSobyanin2010a}%
  \BibitemOpen
  \bibfield  {author} {\bibinfo {author} {\bibfnamefont {{\relax Ya}.~N.}\
  \bibnamefont {Istomin}}\ and\ \bibinfo {author} {\bibfnamefont {D.~N.}\
  \bibnamefont {Sob'yanin}},\ }\href@noop {} {\bibfield  {journal} {\bibinfo
  {journal} {Astron. Rep.}\ }\textbf {\bibinfo {volume} {54}},\ \bibinfo
  {pages} {338} (\bibinfo {year} {2010}{\natexlab{a}})}\BibitemShut {NoStop}%
\bibitem [{\citenamefont {Istomin}\ and\ \citenamefont
  {Sob'yanin}(2010{\natexlab{b}})}]{IstominSobyanin2010b}%
  \BibitemOpen
  \bibfield  {author} {\bibinfo {author} {\bibfnamefont {{\relax Ya}.~N.}\
  \bibnamefont {Istomin}}\ and\ \bibinfo {author} {\bibfnamefont {D.~N.}\
  \bibnamefont {Sob'yanin}},\ }\href@noop {} {\bibfield  {journal} {\bibinfo
  {journal} {Astron. Rep.}\ }\textbf {\bibinfo {volume} {54}},\ \bibinfo
  {pages} {355} (\bibinfo {year} {2010}{\natexlab{b}})}\BibitemShut {NoStop}%
\bibitem [{\citenamefont {Gurevich}\ and\ \citenamefont
  {Istomin}(1985)}]{Gurevich1985}%
  \BibitemOpen
  \bibfield  {author} {\bibinfo {author} {\bibfnamefont {A.~V.}\ \bibnamefont
  {Gurevich}}\ and\ \bibinfo {author} {\bibfnamefont {{\relax Ya}.~N.}\
  \bibnamefont {Istomin}},\ }\href@noop {} {\bibfield  {journal} {\bibinfo
  {journal} {Sov. Phys. JETP}\ }\textbf {\bibinfo {volume} {62}},\ \bibinfo
  {pages} {1} (\bibinfo {year} {1985})}\BibitemShut {NoStop}%
\bibitem [{\citenamefont {Beskin}\ \emph {et~al.}(1993)\citenamefont {Beskin},
  \citenamefont {Gurevich},\ and\ \citenamefont {\relax{Ya}.
  N.~Istomin}}]{BGI1993}%
  \BibitemOpen
  \bibfield  {author} {\bibinfo {author} {\bibfnamefont {V.~S.}\ \bibnamefont
  {Beskin}}, \bibinfo {author} {\bibfnamefont {A.~V.}\ \bibnamefont
  {Gurevich}}, \ and\ \bibinfo {author} {\bibnamefont {\relax{Ya}.
  N.~Istomin}},\ }\href@noop {} {\emph {\bibinfo {title} {Physics of the Pulsar
  Magnetosphere}}}\ (\bibinfo  {publisher} {Cambridge University Press},\
  \bibinfo {address} {Cambridge},\ \bibinfo {year} {1993})\ p.\ \bibinfo
  {pages} {193}\BibitemShut {NoStop}%
\bibitem [{\citenamefont {Deutsch}(1955)}]{Deutsch1955}%
  \BibitemOpen
  \bibfield  {author} {\bibinfo {author} {\bibfnamefont {A.~J.}\ \bibnamefont
  {Deutsch}},\ }\href@noop {} {\bibfield  {journal} {\bibinfo  {journal} {Ann.
  Astrophys.}\ }\textbf {\bibinfo {volume} {18}},\ \bibinfo {pages} {1}
  (\bibinfo {year} {1955})}\BibitemShut {NoStop}%
\bibitem [{\citenamefont {Shukre}\ and\ \citenamefont
  {Radhakrishnan}(1982)}]{ShukreRadhakrishnan1982}%
  \BibitemOpen
  \bibfield  {author} {\bibinfo {author} {\bibfnamefont {C.~S.}\ \bibnamefont
  {Shukre}}\ and\ \bibinfo {author} {\bibfnamefont {V.}~\bibnamefont
  {Radhakrishnan}},\ }\href@noop {} {\bibfield  {journal} {\bibinfo  {journal}
  {Astrophys. J.}\ }\textbf {\bibinfo {volume} {258}},\ \bibinfo {pages} {121}
  (\bibinfo {year} {1982})}\BibitemShut {NoStop}%
\bibitem [{\citenamefont {Ruderman}\ and\ \citenamefont
  {Sutherland}(1975)}]{RudermanSutherland1975}%
  \BibitemOpen
  \bibfield  {author} {\bibinfo {author} {\bibfnamefont {M.~A.}\ \bibnamefont
  {Ruderman}}\ and\ \bibinfo {author} {\bibfnamefont {P.~G.}\ \bibnamefont
  {Sutherland}},\ }\href@noop {} {\bibfield  {journal} {\bibinfo  {journal}
  {Astrophys. J.}\ }\textbf {\bibinfo {volume} {196}},\ \bibinfo {pages} {51}
  (\bibinfo {year} {1975})}\BibitemShut {NoStop}%
\bibitem [{\citenamefont {Barsukov}\ \emph {et~al.}(2007)\citenamefont
  {Barsukov}, \citenamefont {Kantor},\ and\ \citenamefont
  {Tsygan}}]{BarsukovKantorTsygan2007}%
  \BibitemOpen
  \bibfield  {author} {\bibinfo {author} {\bibfnamefont {D.~P.}\ \bibnamefont
  {Barsukov}}, \bibinfo {author} {\bibfnamefont {E.~M.}\ \bibnamefont
  {Kantor}}, \ and\ \bibinfo {author} {\bibfnamefont {A.~I.}\ \bibnamefont
  {Tsygan}},\ }\href@noop {} {\bibfield  {journal} {\bibinfo  {journal}
  {Astron. Rep.}\ }\textbf {\bibinfo {volume} {51}},\ \bibinfo {pages} {469}
  (\bibinfo {year} {2007})}\BibitemShut {NoStop}%
\bibitem [{\citenamefont {Barsukov}\ \emph {et~al.}(2009)\citenamefont
  {Barsukov}, \citenamefont {Polyakova},\ and\ \citenamefont
  {Tsygan}}]{BarsukovPolyakovaTsygan2009}%
  \BibitemOpen
  \bibfield  {author} {\bibinfo {author} {\bibfnamefont {D.~P.}\ \bibnamefont
  {Barsukov}}, \bibinfo {author} {\bibfnamefont {P.~I.}\ \bibnamefont
  {Polyakova}}, \ and\ \bibinfo {author} {\bibfnamefont {A.~I.}\ \bibnamefont
  {Tsygan}},\ }\href@noop {} {\bibfield  {journal} {\bibinfo  {journal}
  {Astron. Rep.}\ }\textbf {\bibinfo {volume} {53}},\ \bibinfo {pages} {86}
  (\bibinfo {year} {2009})}\BibitemShut {NoStop}%
\bibitem [{\citenamefont {Istomin}\ and\ \citenamefont
  {Sobyanin}(2007)}]{IstominSobyanin2007}%
  \BibitemOpen
  \bibfield  {author} {\bibinfo {author} {\bibfnamefont {{\relax Ya}.~N.}\
  \bibnamefont {Istomin}}\ and\ \bibinfo {author} {\bibfnamefont {D.~N.}\
  \bibnamefont {Sobyanin}},\ }\href@noop {} {\bibfield  {journal} {\bibinfo
  {journal} {Astron. Lett.}\ }\textbf {\bibinfo {volume} {33}},\ \bibinfo
  {pages} {660} (\bibinfo {year} {2007})}\BibitemShut {NoStop}%
\bibitem [{\citenamefont {Beskin}(1982)}]{Beskin1982}%
  \BibitemOpen
  \bibfield  {author} {\bibinfo {author} {\bibfnamefont {V.~S.}\ \bibnamefont
  {Beskin}},\ }\href@noop {} {\bibfield  {journal} {\bibinfo  {journal}
  {Astrofizika}\ }\textbf {\bibinfo {volume} {18}},\ \bibinfo {pages} {439}
  (\bibinfo {year} {1982})}\BibitemShut {NoStop}%
\bibitem [{\citenamefont {Sokolov}\ and\ \citenamefont
  {Ternov}()}]{SokolovTernov1974}%
  \BibitemOpen
  \bibfield  {author} {\bibinfo {author} {\bibfnamefont {A.~A.}\ \bibnamefont
  {Sokolov}}\ and\ \bibinfo {author} {\bibfnamefont {I.~M.}\ \bibnamefont
  {Ternov}},\ }\href@noop {} {\emph {\bibinfo {title} {Synchrotron Radiation
  from Relativistic Electrons}}},\ \bibinfo {note} {(Nauka, Moscow, 1974,
  p.~127; American Institute of Physics, New York, 1986)}\BibitemShut {NoStop}%
\bibitem [{\citenamefont {Gradshteyn}\ and\ \citenamefont
  {Ryzhik}()}]{GradsteinRyzhik1963}%
  \BibitemOpen
  \bibfield  {author} {\bibinfo {author} {\bibfnamefont {I.~S.}\ \bibnamefont
  {Gradshteyn}}\ and\ \bibinfo {author} {\bibfnamefont {I.~M.}\ \bibnamefont
  {Ryzhik}},\ }\href@noop {} {\emph {\bibinfo {title} {Table of Integrals,
  Series and Products}}},\ \bibinfo {note} {(Fizmatlit, Moscow, 1963, p.~698;
  Academic, New York, 1980)}\BibitemShut {NoStop}%
\bibitem [{\citenamefont {Nikiforov}\ and\ \citenamefont
  {Uvarov}()}]{NikiforovUvarov}%
  \BibitemOpen
  \bibfield  {author} {\bibinfo {author} {\bibfnamefont {A.~F.}\ \bibnamefont
  {Nikiforov}}\ and\ \bibinfo {author} {\bibfnamefont {V.~B.}\ \bibnamefont
  {Uvarov}},\ }\href@noop {} {\emph {\bibinfo {title} {Special Functions of
  Mathematical Physics}}},\ \bibinfo {note} {(Nauka, Moscow, 1984, p.~180;
  Birkhauser, Basel, Switzerland, 1988)}\BibitemShut {NoStop}%
\bibitem [{\citenamefont {Prudnikov}\ \emph {et~al.}(1989)\citenamefont
  {Prudnikov}, \citenamefont {\relax{Yu}. A.~Brychkov},\ and\ \citenamefont
  {Marichev}}]{PrudnikovBrychkovMarichev1989}%
  \BibitemOpen
  \bibfield  {author} {\bibinfo {author} {\bibfnamefont {A.~P.}\ \bibnamefont
  {Prudnikov}}, \bibinfo {author} {\bibnamefont {\relax{Yu}. A.~Brychkov}}, \
  and\ \bibinfo {author} {\bibfnamefont {O.~I.}\ \bibnamefont {Marichev}},\
  }\href@noop {} {\bibfield  {journal} {\bibinfo  {journal} {Itogi Nauki Tekh.,
  Ser.: Mat. Anal.}\ }\textbf {\bibinfo {volume} {27}},\ \bibinfo {pages} {3}
  (\bibinfo {year} {1989})},\ \translation{{J. Sov. Math.} \textbf{54}, 1239
  (1991)}\BibitemShut {NoStop}%
\bibitem [{\citenamefont {\relax{Yu}. A.~Brychkov}\ and\ \citenamefont
  {Prudnikov}()}]{BrychkovPrudnikov1977}%
  \BibitemOpen
  \bibfield  {author} {\bibinfo {author} {\bibnamefont {\relax{Yu}.
  A.~Brychkov}}\ and\ \bibinfo {author} {\bibfnamefont {A.~P.}\ \bibnamefont
  {Prudnikov}},\ }\href@noop {} {\emph {\bibinfo {title} {Integral
  Transformations of Generalized Functions}}},\ \bibinfo {note} {(Nauka,
  Moscow, 1977, p.~44; Gordon and Breach, New York, 1989)}\BibitemShut
  {NoStop}%
\bibitem [{\citenamefont {Sob'yanin}(2010)}]{SobyaninDissertation}%
  \BibitemOpen
  \bibfield  {author} {\bibinfo {author} {\bibfnamefont {D.~N.}\ \bibnamefont
  {Sob'yanin}},\ }\href@noop {} {\bibinfo {type} {\relax{Candidate's
  Dissertation (Phys.--Math.)}}},\ \bibinfo  {school} {Moscow Institute of
  Physics and Technology}, \bibinfo {address} {Moscow} (\bibinfo {year}
  {2010})\BibitemShut {NoStop}%
\bibitem [{\citenamefont {Pinney}()}]{Pinney1961}%
  \BibitemOpen
  \bibfield  {author} {\bibinfo {author} {\bibfnamefont {E.}~\bibnamefont
  {Pinney}},\ }\href@noop {} {\emph {\bibinfo {title} {Ordinary
  Difference-Differential Equations}}},\ \bibinfo {note} {(University of
  California Press, Berkeley, California, 1958; Inostrannaya Literatura,
  Moscow, 1961)}\BibitemShut {NoStop}%
\bibitem [{\citenamefont {Fedoryuk}(1977)}]{Fedoryuk1977}%
  \BibitemOpen
  \bibfield  {author} {\bibinfo {author} {\bibfnamefont {M.~V.}\ \bibnamefont
  {Fedoryuk}},\ }\href@noop {} {\emph {\bibinfo {title} {Method of Steepest
  Descent}}}\ (\bibinfo  {publisher} {Nauka},\ \bibinfo {address} {Moscow},\
  \bibinfo {year} {1977})\ p.~\bibinfo {pages} {29},\ \bibinfo {note} {[in
  Russian]}\BibitemShut {NoStop}%
\end{thebibliography}%
\begin{flushright}
\textit{Translated by V. Astakhov}
\end{flushright}
\newpage
\begin{figure}[h]
\centering
\begin{overpic}[angle=-90]{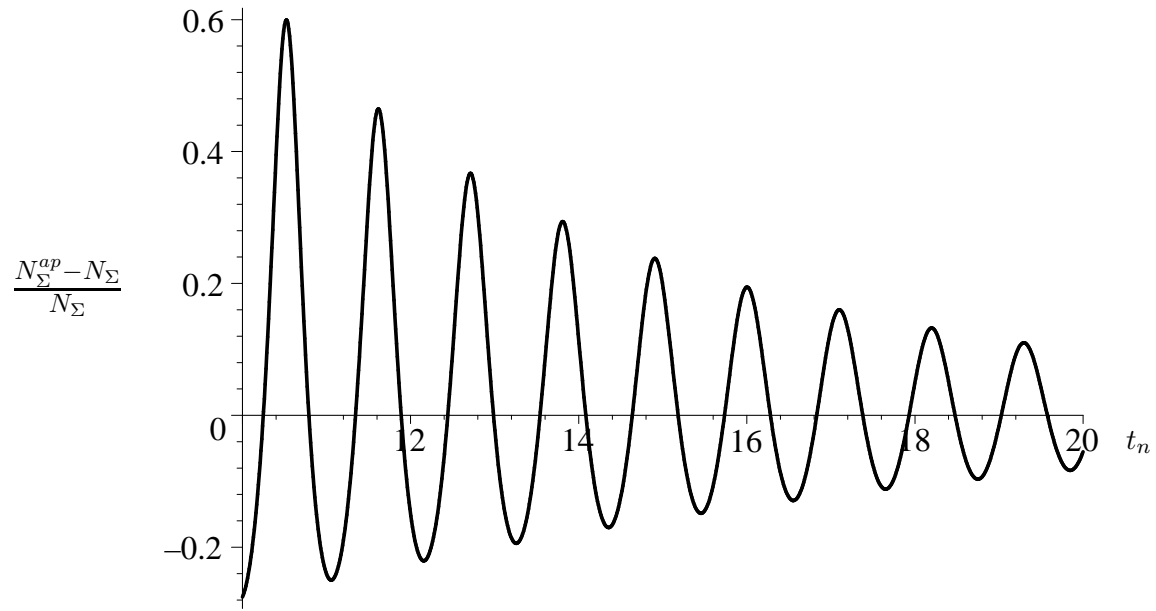}
\put(103,17){$t_n$}
\put(-16,33){\large$\frac{N^{ap}_\Sigma-N_\Sigma}{N_\Sigma}$}
\end{overpic}
\caption{Accuracy of the approximation $N^{ap}_\Sigma$ of the number of electron-positron pairs $N_\Sigma$ versus normalized time $t_n=t/\tau$ for $N_\tau=10^5$.}
\end{figure}
\end{document}